\documentclass[prd,reprint, nofootinbib,amsmath,amssymb, aps,superscriptaddress, floatfix]{revtex4-1}

\usepackage{graphicx}
\usepackage{color}
\usepackage{braket}
\usepackage{cancel}
\usepackage{mathrsfs}
\usepackage{enumitem}
\usepackage{amsthm}
\usepackage{mathtools}
\usepackage[normalem]{ulem}

\graphicspath{{./figs/}}

\theoremstyle{remark}

\usepackage{bbold}
\usepackage{comment}

\definecolor{darkgreen}{rgb}{0,0.5,0}
\definecolor{darkblue}{rgb}{0,0,0.6}
\definecolor{purple}{rgb}{0.4,.2,0.7}
\usepackage[colorlinks=true,citecolor=darkgreen,linkcolor=black,urlcolor=purple]{hyperref}
\usepackage[export]{adjustbox}

\DeclareMathOperator{\Tr}{Tr}
\def\be{\begin{eqnarray}}
\def\ee{\end{eqnarray}}

\newcommand{\bea}{\begin{eqnarray}}
\newcommand{\eea}{\end{eqnarray}}
\def\ben{\begin{equation}}
\def\een{\end{equation}}

\let\z=\zeta   
     \let\r=v

\def\nn{\nonumber}

\def\be{\begin{equation}}
\def\ee{\end{equation}}
\def\ba{\begin{array}}
\def\ea{\end{array}}

\def\mO{{\cal O}}

\def\ba#1\ea{\begin{align}#1\end{align}}
\def\bs#1\es{\begin{split}#1\end{split}}

\allowdisplaybreaks  

\def \be {\begin{equation}}
\def \ee {\end{equation}}
	
\begin{document}
\title{Continuing past the inner horizon using WKB}
\author{Shadi Ali Ahmad}
\affiliation{Center for Cosmology and Particle Physics, New York University, New York, NY 10003, USA }
\author{Ahmed Almheiri}
\affiliation{Center for Cosmology and Particle Physics, New York University, New York, NY 10003, USA }
\affiliation{New York University Abu Dhabi, Abu Dhabi, P.O. Box 129188, United Arab Emirates, }
\affiliation{Institute for Advanced Study, Princeton, NJ 08540, USA }
\author{Simon Lin}
\affiliation{New York University Abu Dhabi, Abu Dhabi, P.O. Box 129188, United Arab Emirates, }

\begin{abstract}

Features of the black hole interior can be extracted from the analytic structure of boundary correlation functions. Working in the geodesic approximation, we find analytic continuations that probe the interior of rotating and charged black holes. These generate contributions from timelike geodesics that thread the interior and emerge in a future universe. We implement these continuations on the momentum space two-point function and exemplify this in several black hole backgrounds. We also identify position space analytic continuations achieving the same task that incorporate different  continued momentum space correlators. These correspond to non-perturbative corrections to the WKB approximation. We demonstrate this explicitly in the rotating BTZ black hole by showing that the interior geodesics contribute to the continued position space correlator and motivate a picture for how these contributions arise in higher dimensions. For AdS Schwarzschild, we identify the analytically continued solution that captures the bouncing geodesic. We discuss the possibility of using these continuations to probe the instability of inner horizons from the boundary. 

\end{abstract}
\maketitle

\section{To see a world in a two-point function} 

Within the analytic structure of boundary correlation functions in AdS/CFT, there are signatures of the entire bulk geometry, including black hole singularities. The history of using correlation functions to study the bulk begins in the early days of AdS/CFT as an application of the extrapolate dictionary \cite{Gubser:1998bc,Witten:1998qj}. Shortly after, it was shown that analytic extensions of correlators in the AdS-Schwarzschild spacetime reveal contributions from null geodesics bouncing off the singularity \cite{Fidkowski:2003nf}. This discovery was elaborated on further in many references including \cite{Festuccia:2005pi,Festuccia:2008zx,Ceplak:2024bja,Dodelson:2023nnr,Dodelson:2023vrw,Dodelson:2024atp,Afkhami-Jeddi:2025wra}.

The utility of this probe goes in both directions of the duality; signatures of gravitational bulk physics can serve as criteria for identifying examples of holographic quantum systems, and corrections to boundary computations can shed light on novel bulk quantum gravity effects. An example of the latter would be investigating possible modifications to the interior of black holes. While there are no obvious modifications expected for AdS-Schwarzschild, the status of the inner horizon in charged and rotating black holes remains in question since the quantum stress tensor diverges near the inner horizon \cite{Hollands:2019whz}. This is expected to induce large backreaction to which boundary correlation functions are plausibly sensitive.

\begin{figure}[t]
 \includegraphics[scale=0.3, valign = c]{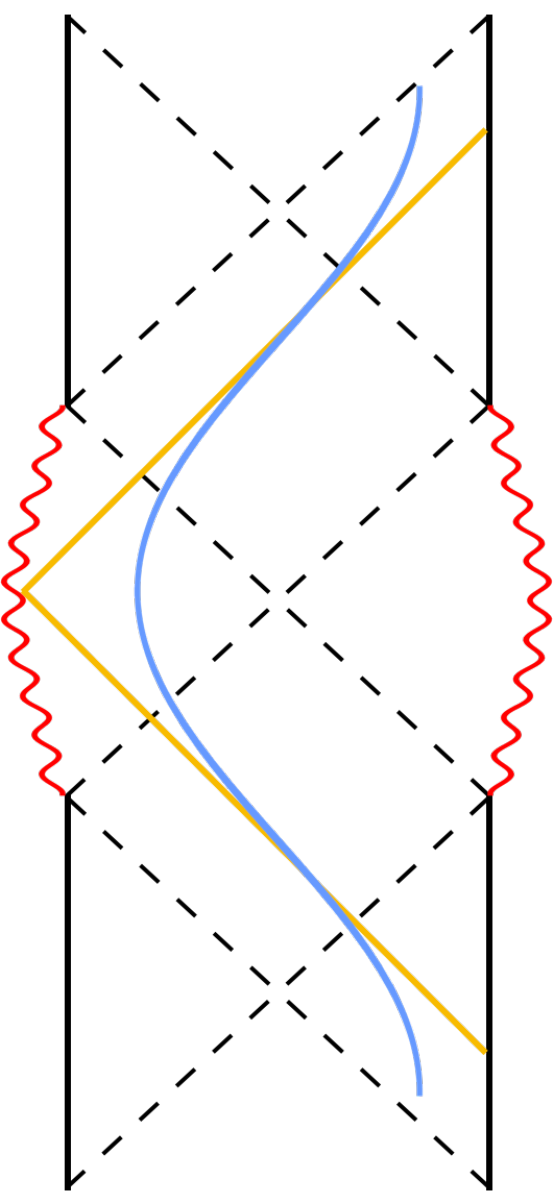}
\caption{\label{RNIntro}
{\footnotesize Shown here are some geodesics that probe behind the inner horizon of a charged or rotating black hole in the maximally extended spacetime. These geodesics cannot reemerge in the same universe.}}
\end{figure}

An immediate obstacle to probing black holes with inner horizons is the absence of structure in the interior that can reflect geodesics back out into the original asymptotic region;\footnote{This refers to a pair of spacelike related asymptotic regions.} see FIG.~\ref{RNIntro}. Furthermore, spacelike geodesics stay away from the inner horizon and accumulate on a maximal volume slice \cite{Hartman:2013qma}.

The way around this obstacle is to consider geodesics that connect timelike related universes in the maximally extended spacetime, see FIG.~\ref{RNIntro}.  This opens up large families of timelike and null geodesics that thread the inner horizon, as well as null geodesics that reflect off the timelike singularity; for example, see \cite{Cruz:1994ir,Gonzalez:2020zfd,Gonzalez:2023jhx,Delsate:2015ina,Grunau:2017uzf} for analysis of such geodesics in various charged and rotating black holes. What remains is to show that these geodesics contribute to boundary correlation functions.

In this paper, we show how to extract contributions of such paths to boundary correlation functions. By working in a large-mass WKB approximation in the bulk, we provide a systematic way of generating these via analytic continuations in both position and momentum space. These continuations have the effect of placing one of the operator insertions in a future universe. For special kinematics, this generates a phase corresponding to a timelike excursion through the interior of the black hole. We should note that this problem has been analyzed previously for the BTZ black hole in many works including \cite{Liu:2002kb,Kraus:2002iv,Hemming:2002kd,Kraus:2002iv,Balasubramanian:2004zu,Balasubramanian:2019qwk} by leveraging the fact that BTZ is a quotient of global AdS. Our analysis goes beyond this and applies to more general black holes.

The outline of this paper is as follows. We begin with a lightning review of the WKB approximation in solving the large-mass limit of bulk wave equations. As a first example, we consider empty AdS in arbitrary dimensions as a simplified setting to define our analytic continuation procedure. We follow this up with applications to black holes with timelike singularities in various dimensions. We also consider black holes without inner horizons in which the analytic continuation generates complex excursions that limit to real bouncing geodesics. 

\textbf{Note added:} The preprints \cite{Ceplak:2025dds, Dodelson:2025jff} appeared while this project was still underway.

\section{\textbf{Geodesics from WKB}}
We start with a lighting review of the WKB method for computing bulk correlation functions in the large-mass limit. We also give an expanded overview in Appendix \ref{app: wkb}. See for example \cite{Festuccia:2005pi} for more details on this method.

We are interested in the boundary Wightman function $G^+(x) \equiv \Tr [e^{-\beta H}\mO(\tau, x)\mO(0)]$ and its analytic continuation in the complex $\tau$ plane. We consider a bulk field with large mass $m$, relative to the AdS scale, on a general asymptotically AdS space. Working in momentum space, the wavefunctions in the Hartle-Hawking state can be solved using WKB perturbatively in $1/m$.  The correlator is then constructed from these wavefunctions and takes the general form
\begin{align}
    \widetilde{G}^+(w,k) \sim e^{m Z(w,k)}+  \ ... , \label{eqn:wkbansatz}
\end{align}
where the bulk spacetime was assumed to have at least two Killing directions and $w,k$ are their associated conserved charges, which we refer to as frequency and momentum respectively, and the ellipses denote perturbative and non-perturbative corrections to the correlator.

The position space correlator is obtained by a Fourier transform
\begin{align}
    G^+(\tau, x) & \sim \int dw dk e^{i w \tau + i k x } e^{mZ}. \label{GE}
\end{align}
where $\tau, x$ are the separations between the operators. We will evaluate such integrals in the semi-classical limit using saddle point methods. These are governed by the equations
\begin{align}
    i \tau = -m \partial_w Z, \quad i x = -m \partial_k Z
\end{align}
The solution is related to the geodesic distance between the insertions through
\begin{align}
    mZ = - i w \tau - i k x - m D,
\end{align}
where $D(\tau,x)$ is the renormalized geodesic length. Plugging this back into the integrand gives the position space correlator in the geodesic approximation
\begin{align}
    G^+(\tau, x) &\sim e^{-m D(\tau, x)}.
\end{align}

We will be interested in continuations of these expressions in both position and momentum space. The goal of this paper is to understand the role of each in determining the contribution from interior geodesics.

\section{\textbf{Lorentzian excursions from analytic continuations}}
In this section, we show how analytic continuations of Euclidean correlators can generate \textit{timelike} contributions. We will first work directly with geodesics and demonstrate that continuations of their conserved charges around branch points generate timelike segments and shifts of the boundary endpoints. We then consider the Fourier transform of momentum space correlations functions for arbitrary boundary endpoints, and show how the continuation of those endpoints deforms the integration contour to pick up the timelike contributions. We warm up with empty AdS here before jumping into black holes.

On Euclidean Poincare AdS with the metric
\begin{align}
    ds^2 = {d\tau^2 + dz^2 + d\vec{x}^2 \over z^2},
\end{align}
the geodesic equation reads
\begin{align}
    z^2 \dot{z}^2 = 1 - z^2(E^2 + L^2).
\end{align}
The parameters $E$ and $L$ are the conserved charges along the geodesic associated to Killing vector fields $(\partial_\tau)^{\mu}$ and $(\partial_x)^{\mu}$ where $x$ is a direction along which we choose to extend the geodesic.\footnote{This discussion applies to any spacetime dimension. Here, we are considering motion along a three dimensional submanifold.} The ``$\, \dot{\ }\, $'' is the proper length derivative.

The equation $\dot{z} = 0$ has two roots corresponding to two turning points of the geodesic $z = \pm z_t \equiv \pm 1/\sqrt{E^2 + L^2}$. The positive branch corresponds to the usual turning point of a semi-circle extending in the upper-half-plane. The negative branch $-z_t$ is not a point in the original spacetime, but a point in the neighboring Poincare patch in the Lorentzian section. A geodesic with the $-z_t$ turning point exits the Euclidean section along the Euclidean time-reflection slice:
\begin{align}
    \includegraphics[scale=.3, valign = c]{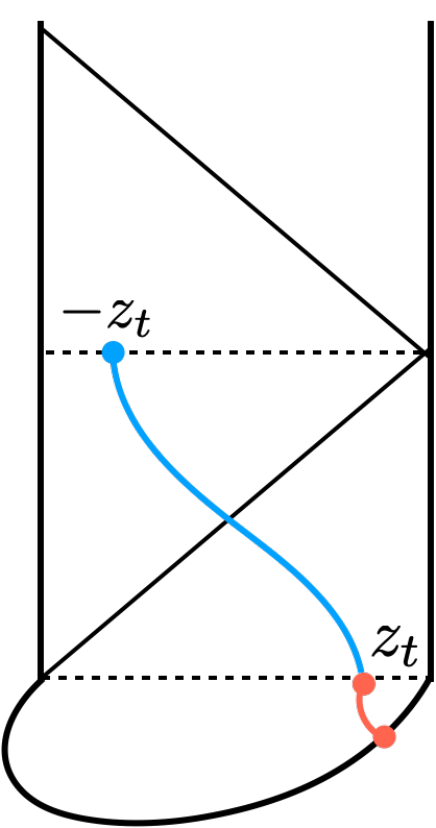}
\end{align}
Our task is to understand the manipulations that exchange the turning points and result in such timelike excursions.

The geodesic length in empty AdS in the Poincare patch is given by the integral expression\footnote{This integral is divergent and requires a renormalization which we have made implicit here. The same applies to all length and WKB action integrals we will consider in this paper.}

\begin{align}
    D &= \int_{\mathcal{C}}{dz \over z \sqrt{1 - z^2( E^2 +  L^2)}}, \\[5pt]
    &=\includegraphics[scale=0.3, valign = c]{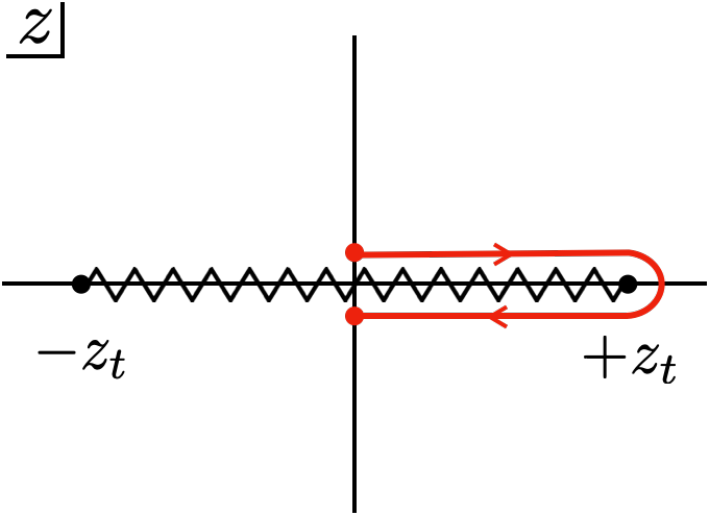}.
\end{align}
 On the second line, we show the integration contour on the $z$ plane; it starts and ends on the boundary after winding around the turning point.\footnote{Other works usually consider one half of this contour going from the boundary to the turning point and multiply by 2.}

This geodesic can be extended to the other turning point $- z_t$ by a continuation in momentum space that swaps the two turning points; we need to rotate $E^2 + L^2$ by $2\pi$ around the origin. We do this by writing  $E^2 + L^2 = (E - i L)(E + i L)$, and winding $E + i L$ anticlockwise around zero. This induces a clockwise rotation of the branch points around the origin by an angle $\pi$. As a result, the integration contour gets dragged to
\begin{align}
		\includegraphics[scale=.3, valign = c]{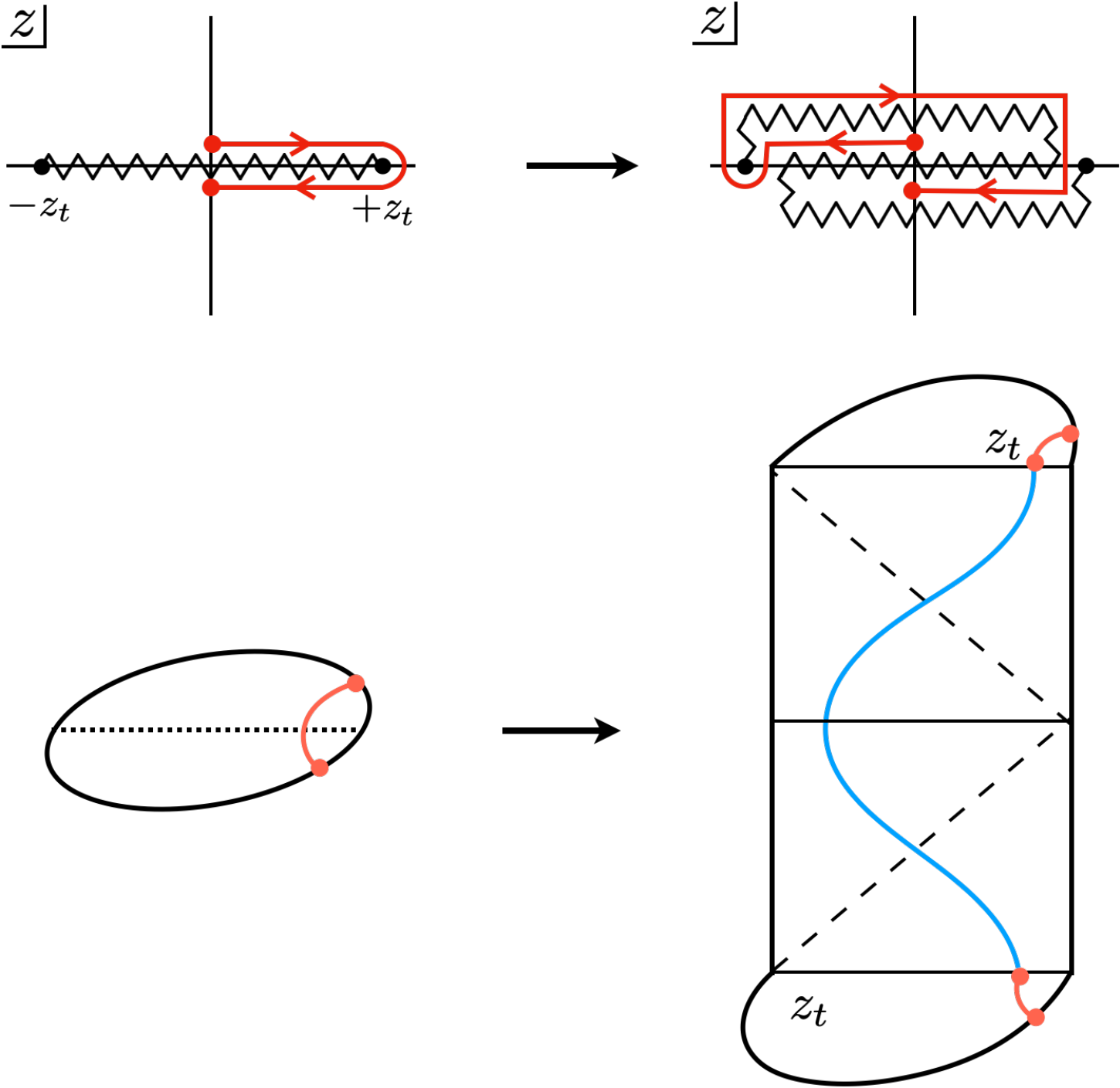}  \nn
\end{align}
The dragged contour cannot be unwound despite $E^2 + L^2$ going back to itself. Then, the geodesic length is shifted $D \to D + 2 \pi i$, where $2 \pi i$ comes from the two contour segments that wind around the origin, one contributing $3\pi i/2$ and the other $\pi i /2$. This imaginary shift is the timelike length of the geodesic in the Lorentzian section.

In addition to changing the length, the separation between the boundary endpoints is also modified by the continuation in $E,L$. From the geodesic conservation equations $E  = \dot{\tau}/z^2, L = \dot{x}/z^2$ we have
\begin{align}
\tau \pm ix &= (E \pm iL) \int_{\mathcal{C}}{z dz \over \sqrt{1 - z^2( E^2 +  L^2)} }, \\
    &= { 1 \over E \mp iL} \int_{\mathcal{C}}{z dz \over \sqrt{1 - z^2} },
\end{align}
where the dependence on $E,L$ was factored out using $z \to z/\sqrt{E^2 + L^2}$. Then, for every $2\pi$ rotation in $E + i L$, we have
\begin{align}
    \tau - i x \to e^{- 2\pi i}(\tau - i x), \quad \tau + i x \to \tau + i x. \label{PoincareTransformation}
\end{align}

We could also consider a $\pi$ rotation that sends $\tau - i x \to e^{- \pi i}(\tau - i x)$ and moves the turning point to the imaginary axis. This coordinate shift places one insertion in the neighboring Poincare patch. This behavior was previously found in \cite{Kravchuk:2018htv} using the action of the conformal group on AdS space. Here we interpret the phase, which also appeared in their work, in terms of the timelike length between the separate patches.

We now connect this back to the WKB method. The output of the WKB analysis is the momentum space correlator given by $e^{mZ}$ where
\begin{gather}
    Z = -\int_{\mathcal{C}}{\sqrt{1 + z^2( w^2 +  k^2)/m^2}dz \over z} .
\end{gather}
where we substituted $E \to -i w/m, ~ L \to -i k/m$.

We have so far established that this correlator can be continued to pick up timelike contributions, and  predicted the required continuation to do so in position space. The goal now is to check this by applying this continuation to the position space correlator written as a Fourier transform from momentum space
\begin{align}
    G^+(\tau, x) & \sim \int dw dk e^{i w \tau + i k x } e^{mZ}. 
\end{align}
This is simple to analyze in polar coordinates
\begin{gather}
    w = \rho \cos \theta, \quad  k = \rho \sin \theta, \quad \nonumber \\ \tau = \z \cos \varphi, \quad x = \z \sin \varphi,  \nonumber
\end{gather}
where the integral takes the form
\begin{align}
  G_E(\tau, x) =  \int d\theta d\rho \rho e^{i \rho \z \cos[\theta - \varphi] + m \ln \rho^2 + m (\ln4 -2)}.
\end{align}
We evaluate this integral using steepest descent in $\rho$.  For real $\z$, the path of steepest descent in $\rho$ begins at the origin and extends parallel to the imaginary axis. It crosses the saddle point at
\begin{align}
    \rho &= {2i m \over \z \cos[\theta - \varphi]  }={2i m \over \tau \cos[\theta] + x \sin[\theta]  }.
\end{align}

Next, we need to track the saddle point and the integration contour under the continuation \eqref{PoincareTransformation}. Rotating $\tau - i x$ around the origin while keeping $\tau + i x$ fixed can be accomplished through
\begin{gather}
    \tau \to e^{i \alpha/2}(\tau \cos \alpha/2 - x \sin \alpha/2), \\
     x \to e^{i \alpha/2}(\tau \sin \alpha/2 + x \cos \alpha/2).
\end{gather}
This multiplies the saddle point by the overall phase $e^{-i\alpha/2}$, rotating it around the origin in the complex $\rho$ plane. The asymptotic region where the integrand vanishes rotates in the same way, and hence the steepest descent follows suit:
\begin{align}
		\includegraphics[scale=0.33, valign = c]{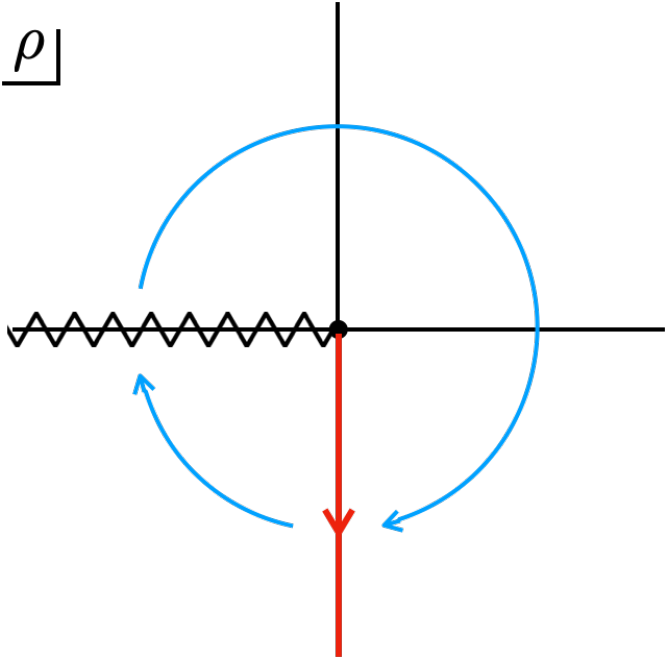}  \nn
\end{align}
This rotation in the presence of the logarithm in the action generates the phase $-i\alpha$. A full clockwise rotation of $\alpha = -2 \pi$ gives the length of the timelike excursion.

We close this section by demonstrating that the timelike contribution can be seen at the level of the evaluated two-point function. For instance, applying the coordinate transformation  \eqref{PoincareTransformation} $k$ times on a 2d CFT two-point function leads to 
\begin{align}
    {1 \over (x - i \tau)^{2h}(x + i \tau)^{2 \bar{h}}} \to  {e^{ 2 \pi i   k \Delta} \over (x - i \tau)^{2h}(x + i \tau)^{2 \bar{h}}},
\end{align}
where $\Delta = h+ \bar{h}$. The extra phase is the conformal dimension of the operator multiplied by the imaginary Lorentzian length of the geodesic as it extends into the future. 

These same techniques apply to global AdS with similar conclusions. Next, we will describe the continuations for black holes, starting with those with inner horizons.
\section{{Through the interior }} 

In this section we apply the above prescription to black holes with inner horizons that connect timelike related asymptotic regions. We find the analytic continuations that generate contributions from excursions through the interior to boundary-anchored correlation functions. 

\subsection{{Rotating BTZ black hole}}  

We begin with the case of a rotating BTZ black hole. The (complex)\footnote{The metric is complex because the Lorentzian metric does not have time reflection symmetry. This choice has the advantage of giving the familiar Lorentzian geodesic equation upon continuation.} Euclidean metric reads~\cite{Banados:1992wn}
\begin{equation}
    ds^2=f(r)d\tau^2+\frac{dr^2}{f(r)}+r^2\left(d\varphi-i \frac{r_+r_-}{r^2}d\tau\right)^2, \label{euclideanBTZ}
\end{equation}
with $f(r)= (r^2-r_+^2)(r^2-r_-^2)/r^2$.
The geodesic equation reads \cite{Cruz:1994ir}
\begin{gather}
	r^2\dot{r}^2 = (r^2 - r_+^2)(r^2 - r_-^2)  - L^2 (r^2 - r_+^2 - r_-^2) \nonumber \\ - E(E r^2 + 2i L r_+ r_-). \label{eq:BTZGeoEq}
\end{gather}
where again $E,L$ are the conserved charges along the geodesics corresponding to $\partial_\tau$ and $\partial_\varphi$. They determine the rate of change of $t, \varphi$ along the geodesic through
\begin{align}
	\dot{\tau} &= {E r^2 + i L r_+ r_- \over (r^2 - r_+^2)(r^2 - r_-^2) }, \\
	\dot{\varphi} &= { L (r^2 - r_-^2 - r_+^2) +i E r_+ r_- \over (r^2 - r_+^2)(r^2 - r_-^2) }.
\end{align}
Let's analyze the turning points as a function of the conserved charges. Setting the right hand side of \eqref{eq:BTZGeoEq} to zero results in a quadratic equation for $r^2$ with roots
\begin{gather}
    2 r_{t_\pm}^2 = E^2 + L^2 + r_+^2 + r_-^2 
     \pm \sqrt{A_+ A_- \bar{A}_+\bar{A}_- }
\end{gather}
where
\begin{align}
    A_\pm &= (E + i L) \pm i (r_+ - r_-), \\ 
    \bar{A}_\pm &= (E - i L) \pm i (r_+ +  r_-).
\end{align}
Thus, the turning point radius has four branch points located at $E = {\textcolor{darkgreen}{\pm}} iL {\textcolor{red}{\pm}}i(r_+ {\textcolor{darkgreen}{\pm}} r_-)$, where different color signs are independent. We see that the turning points are swapped by rotating  any of the $A$'s around zero.

Next we analyze how the geodesic length is affected by swapping the turning points. Consider a geodesic anchored to Euclidean time on the boundary. The geodesic length of such a path is given by the integral expression
\begin{align}
    D &= \int_{\mathcal{C}}{ r dr \over  \sqrt{(r^2 - r_{t_-}^2)(r^2 - r_{t_+}^2) } }, \\
    &= \quad \includegraphics[scale=0.33, valign = c]{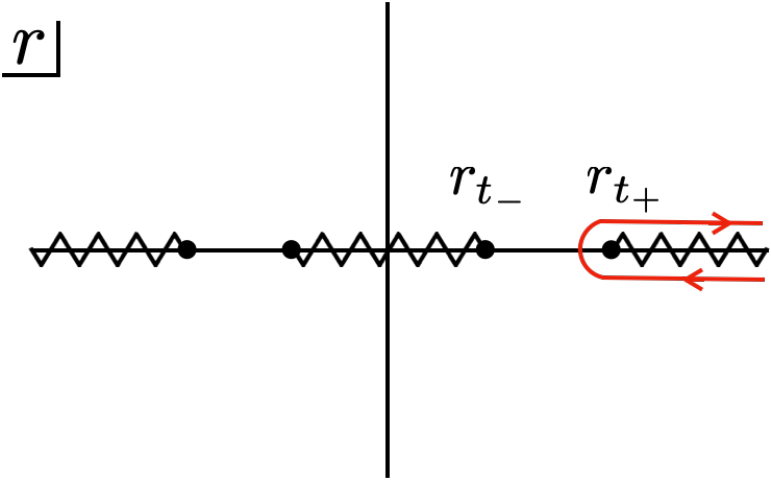}, \\
    &=\ln4-\frac{1}{2}\log(A_+A_-\bar{A}_+\bar{A}_-)
\end{align}
where the contour follows the path of the geodesic to and from $\infty$ and around the outer turning point. The  contour gets dragged when the turning points are swapped. For example, rotating $A_-$ around zero results in
\begin{align}
    \includegraphics[scale=0.33, valign = c]{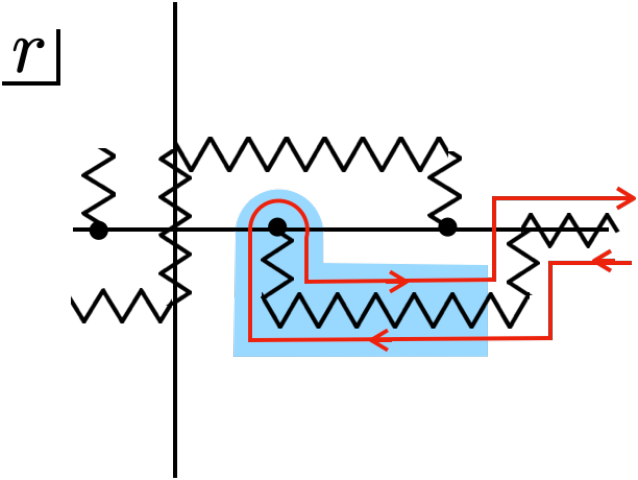}. 
\end{align}
The blue highlighted contour evaluates to $\pi i$ and captures the blue Lorentzian segment:
\begin{align}
    \includegraphics[scale=0.28, valign = c]{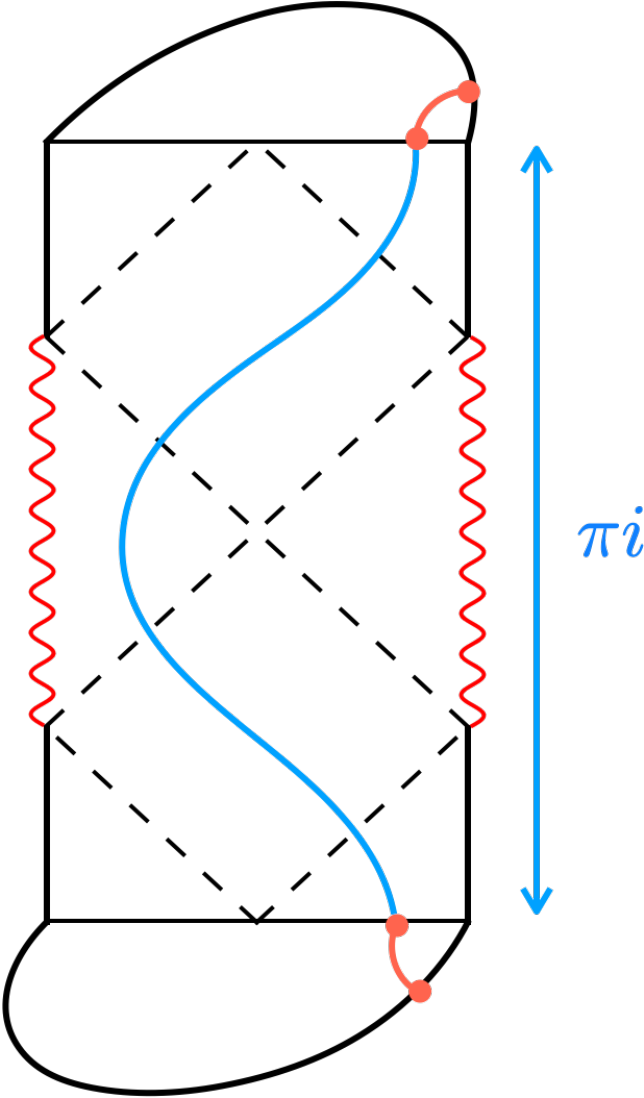}.  \label{eq:BTZtimelike}
\end{align}
This timelike length is independent of the charges because all such geodesics are related by symmetry.

\begin{figure}
    \centering
    \includegraphics[width=\linewidth, valign = c]{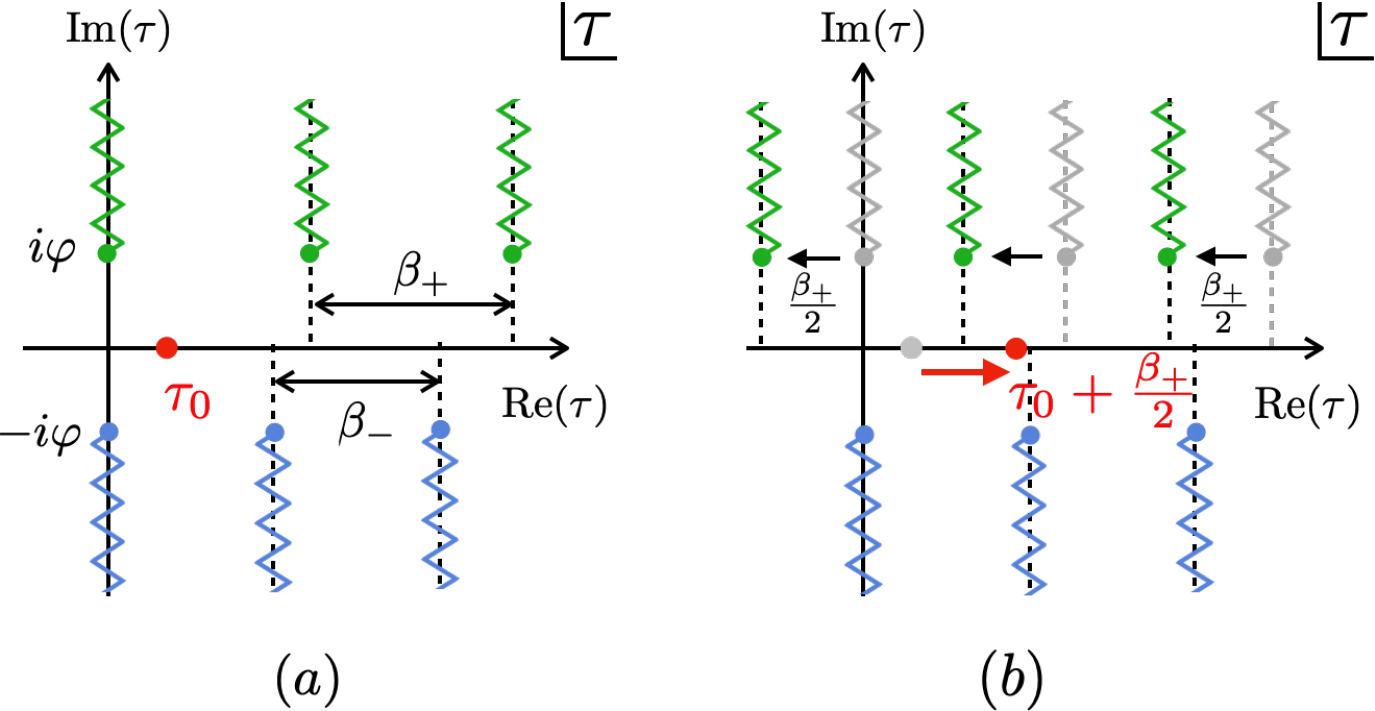}    
    \caption{(a) The analytic structure of the position space correlator $G^+(\tau,\varphi)$. The correlator has branch cuts separated by the chiral temperatures $\{\beta,\bar{\beta}\}$. (b) The analytic continuation that generates the Lorentzian excursion. In the $\tau$ plane, the continuations shifts the branch points of the holomorphic part of $G^+$ (labeled in green) in addition to moving $\tau$. Note that the anti-holomorphic part of $G^+$ is left unchanged.}
    \label{fig:BTZTimeContour}
\end{figure}

To find the induced shifts in the boundary coordinates, we use the shortcut of taking derivatives of the WKB phase
\begin{align}
    Z &= -\int_{\mathcal{C}}{   \sqrt{(r^2 - r_{t_-}^2)(r^2 - r_{t_+}^2)/r^2 }  \over f(r)} dr,\\
     &=  {   A_+\ln {A_+ \over 2} - A_-\ln {A_-\over 2} \over 2 i (r_+ - r_-)} + {   \bar{A}_+\ln {\bar{A}_+\over 2} - \bar{A}_-\ln {\bar{A}_-\over 2} \over 2 i (r_+ + r_-)} \label{eq:ZBTZ} 
\end{align}
Taking derivatives with respect to $E$ and $L$ returns the separations in boundary time and space, respectively, leading to
\begin{align}
    \tau &=   { \beta \over 4 \pi i}\ln {A_+ \over A_-}+ {  \bar{\beta} \over 4 \pi i}  \ln {\bar{A}_+ \over \bar{A}_-} \label{eq:Zequation1}\\
    \varphi &= { \beta \over 4 \pi}\ln {A_+ \over A_-} - {  \bar{\beta} \over 4 \pi} \ln {\bar{A}_+ \over \bar{A}_-}\label{eq:Zequation2}
\end{align}
where $\{\beta,\bar{\beta}\} \equiv \{\frac{2\pi}{r_+-r_-},\frac{2\pi}{r_++r_-}\}$ are the chiral temperatures of the black hole.  We see that a $2 \pi i$ clockwise rotation of $A_-$ shifts $\tau \to \tau + \beta/2$ and $\varphi \to \varphi + i\beta/2$.\footnote{We note that this continuation does not agree with the proposal in \cite{Balasubramanian:2019qwk} for achieving the same task.} To determine the path  this shift must take, we consider the $A$'s
in terms of the coordinates, which read
\begin{align}
    {A_\pm } =  {\pm4 \pi i\over \beta} { 1\over 1 - e^{\mp 2\pi  y/\beta}}, \  {\bar{A}_\pm } =  {\pm4 \pi i\over \bar{\beta}} { 1\over 1 - e^{\pm 2\pi  \bar{y}/\bar{\beta}}}\label{eq:Zequation3}
\end{align}
where $\{y,\bar{y}\}\equiv \{\varphi + i \tau,\varphi-i\tau\}$. We focus on $y$ since we are rotating $A_-$. Since the path that rotates $A_-$ must be able to change its sign, it must pass $y = i \beta n$, for any $n \in \mathbb{Z}$, from the right. In the complex $\tau$ plane, this corresponds to shifting $\tau$ to the right while moving the branchpoints at $i \varphi$ to the left. See FIG.~\eqref{fig:BTZTimeContour}.

All the technology developed so far in this section can be used to construct the momentum space correlation function in the WKB limit.  This generates the correlator with boundary condition at the horizon being both ingoing plus outgoing as well as Dirichlet for the normalizable mode at the asymptotic boundary. It takes the form
\begin{align}
    \tilde{G}^+(w,k) = e^{mZ}\Big|_{E \to - i w/m, \ L \to - i k/m}, \label{eqn:BTZGsemi}
\end{align}
where $w,k$ are the frequency and momentum of the field. It is not hard to check that is agrees with the exact momentum space correlation function
\begin{align}
    \tilde{G}^+(p,\bar{p}) \sim e^{-{\bar{p}\beta - p \bar{\beta} \over 4}}\Gamma\!\left[{m \over 2} \pm {\bar{p} \beta \over 4 \pi i}\right]\Gamma\!\left[{m \over 2} \pm  {p \bar{\beta} \over 4 \pi i}\right] \label{eq:exactBTZ}
\end{align}
by approximating the Gamma functions using Stirling's formula (for arguments with positive real part), where $\{p,\bar{p}\} \equiv  \{k+iw,k-iw\}$. The standard WKB prescription with the exterior turning point gives the correlator that probes the Euclidean geometry, and our continuation in frequency (or energy) modifies it to capture the effect of probing the interior.

It is a separate question entirely whether the timelike excursions actually contribute to the position space correlation function. We investigate this by analyzing the steepest descent contour of the Fourier transform of the momentum space correlator and see whether the contour can be deformed to pick up saddles describing the timelike geodesics. The task will be to track how the steepest descent contour and saddles move as $y_+$ is shifted according to the prescription above.

As before, we consider
\begin{align}
    G^+(\tau, \varphi) = \int dw dk e^{i w \tau + i k \varphi} \tilde{G}^+(w,k).
\end{align}
It will be convenient to work in the complex coordinates $y,\bar{y}$ and to note that the WKB phase splits into $Z = Z(p) +\bar{Z}(\bar{p})$. We focus our analysis on the phase
\begin{align}
    i {\bar{p}\, y \over 2} + m \bar{Z}(\bar{p}).
\end{align}
We take the defining integration contour to be the entire real line.\footnote{This is more clear if we continue to Lorentzian space, replacing $\tau \to i t$ and $w \to i \tilde{w}$.} The saddlepoint equation is
\begin{equation}
\label{eq:BTZsaddle}
    \frac{2\pi y}{\beta} = \ln \frac{A_+}{A_-} \quad \Rightarrow \quad \bar{p} = \frac{2\pi i m}{\beta} \coth\frac{\pi y}{\beta}.
\end{equation}
This describes an infinite number of saddle points distributed among an infinite number of sheets, with at most a single saddle per sheet. We begin with $y_0 = i\tau_0+\epsilon$ which guarantees a saddle on the principal sheet whose steepest descent path can be continuously deformed to the defining integration contour along the real axis.

\begin{figure}
    \centering
    \includegraphics[width=\linewidth]{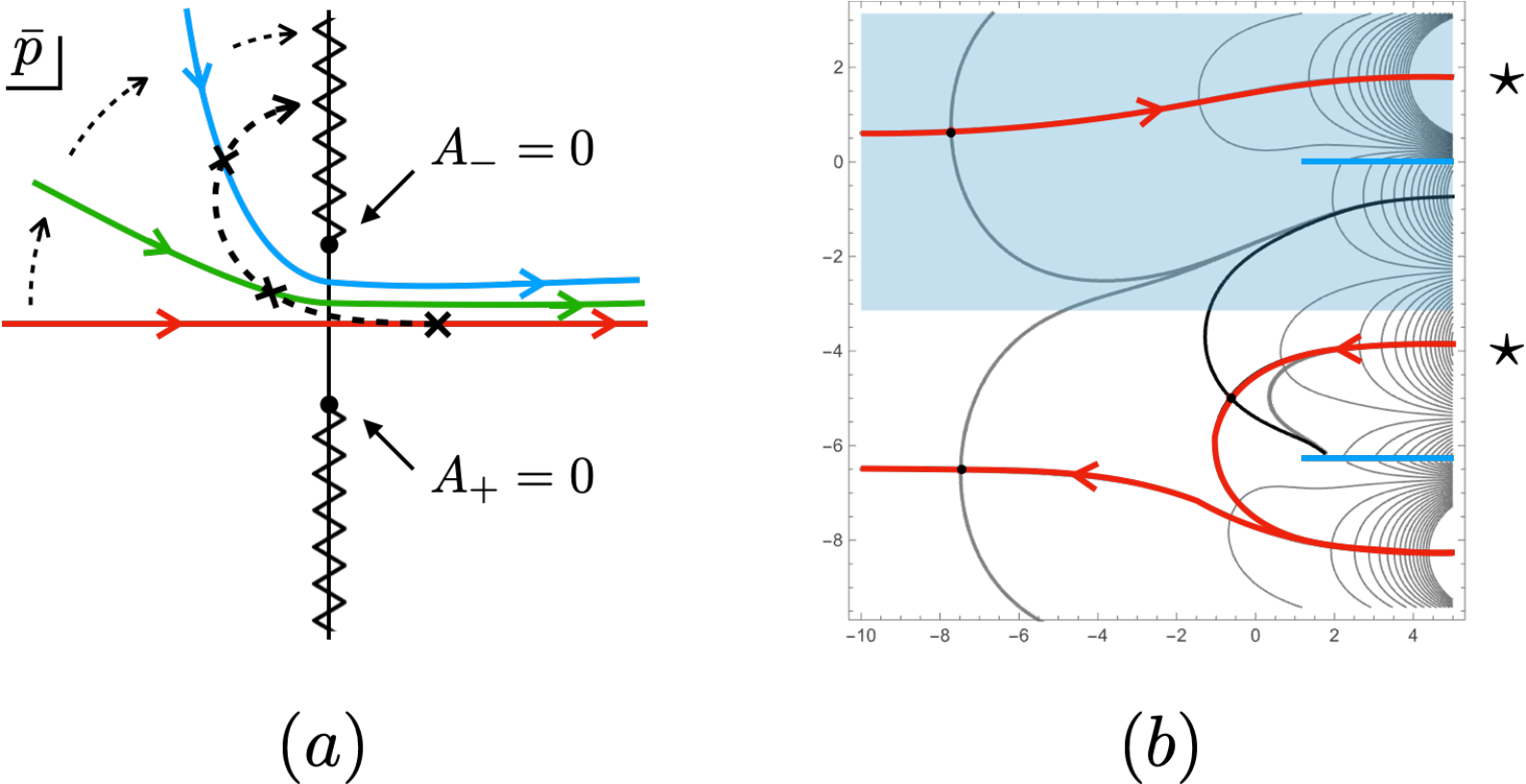}
    \caption{(a) Path of the steepest descent contour in the $\bar{p}$ plane. Under the continuation, the left side of the steepest descent contour peels off of the real axis and bends clockwise through the branch cut of $A_-$. The contour eventually breaks due to a Stokes phenomenon and it becomes unclear in the $\bar{p}$ coordinate if it is deformable to the defining contour. (b) The steepest descent contour in the unwrapped coordinate $\bar{q}$. The principal sheet (a) is mapped to the blue shaded region. This coordinate system makes clear that the steepest descent path (red) continues to be anchored to the asymptotic regions (stars) of the defining contour and can be deformed to it.}
    \label{fig:saddle}
\end{figure}

After a shift $y_0 \to y_0 + i \beta$, the saddle point takes a full turn around the branch point defined by $A_-=0$, stopping where it began but on the second sheet. The steepest descent contour also moves and the asymptotic region of this contour follows the saddle to the next sheet. This is illustrated in FIG.~\ref{fig:saddle}(a).

Before a complete turn of the saddle point, a Stokes phenomenon breaks the path of steepest descent into two segments subtending two sheets, one of which contains the saddle point of interest. This is best seen by working in a uniformizing coordinate $\bar{q}$ in which the branch cut of $\ln A_-$ is unwound; we define the $\bar{q}$ plane through $e^{\bar{q}} \equiv A_-(\bar{p})$, see appendix \ref{app:contour} for more details. In the $\bar{q}$ plane, each sheet of the $\bar{p}$ plane is mapped to an infinite horizontal strip of width $2 \pi i$, each of which contains one instance of the  $\ln A_+$ branch cut. The new contour picks up two additional saddle points originating from the Jacobian factor of the coordinate transformation, but  cancel against each other in the large $m$ limit. The main contribution remains to be the one in \eqref{eq:BTZsaddle}, which simply picks up a phase $\pi i$. This is illustrated in FIG.~\ref{fig:saddle}(b) 

If we continue past $\mathrm{Im}(y) = 3 \beta /2$, we encounter  an additional Stokes phenomenon that removes the principal saddle point from the path of steepest descent homologous to the defining integral. We include a careful analysis of this in appendix \ref{app:contour}. This result suggests that the WKB methods only captures a single excursion into the interior and fails to describe the behavior of the two-point function beyond $\mathrm{Im}(y) = 3 \beta /2$.

A hint of the problem can be seen from comparing the analytic structure of the semi-classical correlator \eqref{eqn:BTZGsemi} to  the  exact answer \eqref{eq:exactBTZ}. The latter only has poles while the former has branch cuts, spoiling the the periodicity around $A_- = 0$. 
The issue is that the semi-classical answer is a good approximation to the exact answer only within a limited domain in the complex $\bar{p}$ plane. This approximation comes from evaluating the integral representation of the Gamma function
\begin{align}
    \Gamma(z) = e^{z \ln z}\int_0^\infty d\chi e^{-z(e^\chi - \chi)}
\end{align}
and via steepest descent methods in the large $z$ limit. When $\mathrm{Arg}[z] \in [-\pi/2,\pi/2]$, a single saddle point lies on the integration contour and captures the semi-classical expression \eqref{eqn:BTZGsemi}. The imaginary $z$ axis is a Stokes line, and crossing it picks up an infinite number of saddles. For example, crossing the $\mathrm{Arg}[z] = - \pi/2$ gives the sum
\begin{align}
    \Gamma(z)  \approx e^{z \ln z - z}\sum_{n = 0}^{\infty} e^{- 2 \pi i z n}.
\end{align}
See \cite{Harlow:2011ny} for more details. Notice that the new terms recover the periodicity by essentially summing over images around the branch point. In our problem $A_- \sim z$, and therefore these extra saddles correspond to geodesics with $n$ timelike excursions of the form shown in \eqref{eq:BTZtimelike}. Note that all of these are independently solutions to the original WKB problem but with different turning points.

Incorporating these non-perturbative effects, the position space correlator takes the form of the Fourier integral
\begin{align}
	G(\tau, \varphi) &= \int \! d\bar{p} \, \mathrm{Exp}\Bigg[i {\bar{p} (y - i \beta n) \over 2}  \nn \\ 
    &+ {m \beta \over 4 \pi i}\left( A_+ \ln A_+ - A_- \ln A_-\right)  - m (n \pi i )\Bigg].
\end{align}
We emphasize that they modify the correlator in two important ways: adding an overall phase of $n \times m \pi i$ and shifting $y$ back by $n$ periods. 

The former reproduces the expected phase from winding geodesics, while the latter changes the location of the saddle to
\begin{align}
    \frac{2\pi }{\beta}\left(y - i \beta n \right) = \ln \frac{A_+}{A_-},
\end{align}
effectively unwinding $A_{-}$ relative to the $n=0$ saddle. This new saddle is then picked up by the steepest descent contour when $\beta<\text{Im}(y) < 5 \beta /2$, for the same reason why the $n=0$ saddle is picked up when $0< \text{Im}(y) < 3 \beta /2$. However, in the overlap region $\beta< \text{Im}(y) < 3 \beta /2$, we encounter the puzzle that the contributions of these saddles differ just by an overall phase. Consequently, neither one dominates over the other and both contribute simultaneously in this regime. We don't have a complete understanding of this yet and will return to it in future work.

As a consistency check of the over all proposal, we apply the continuation on the exact holographic answer at large $c$ in position space. At large $c$, the torus two-point function is dominated by the identity and double-trace blocks that lead to
\begin{align}
    G(\tau, \varphi) = \sum_{k = - \infty}^\infty {1 \over \sinh^{2 h}\left( \pi {\varphi + i \tau +  2 \pi k \over \beta}\right) \sinh^{2 \bar{h}}\left( \pi {   \varphi - i \tau + 2 \pi k\over \bar{\beta}}\right)},
\end{align}
where $k$ indexes the images around the spatial circle. For any $k$, it is simple enough to  check that the above continuation transforms  $G \to G e^{-\Delta i \pi}$, where $\Delta = h + \bar{h}$ (we used the fact that $h - \bar{h} \in \mathbb{Z}$), see FIG.~\ref{fig:BTZTimeContour}. This reproduces the imaginary length $\pi i$. Furthermore, applying the continuation $n$ times gives $n \pi i$.

\subsection{{ AdS Reissner-Nordstrom black hole}}
Next, we analyze the case of AdS Reissner-Nordstrom in five spacetime dimensions. Its Euclidean metric is
\begin{equation}
    ds^2 = f(r)dt^2+\frac{dr^2}{f(r)}+r^2d\Omega_3^2, \label{EuclideanRN}
\end{equation}
where $d\Omega^2_3$ is the round metric on $S^3$ and
\begin{align}
    f(r) &= {(r^2 - r_+^2)(r^2 - r_-^2)(r^2 + 1 + r_+^2 + r_-^2) \over r^4},\\
    &= {r^4 - 2 M r^2 + Q^2 + r^6 \over r^4}.\label{rnf}
\end{align}
Geodesics on this background satisfy\cite{Gonzalez:2020zfd,Gonzalez:2023jhx}
\begin{align}
	\dot{r}^2 + E^2 + L^2 {f(r) \over r^2} =f(r).
\end{align}
The turning point equation $\dot{r} = 0$ is quartic in $r^2$. In the regime with two positive and two conjugate complex roots, the geodesic length is given by
\begin{align}
    D &= \int_{\mathcal{C}}{  dr \over  \sqrt{f(r) \left(1 -L^2 /r^2 \right) - E^2 } }, \\[5pt]
    &= \quad \includegraphics[scale=0.33, valign = c]{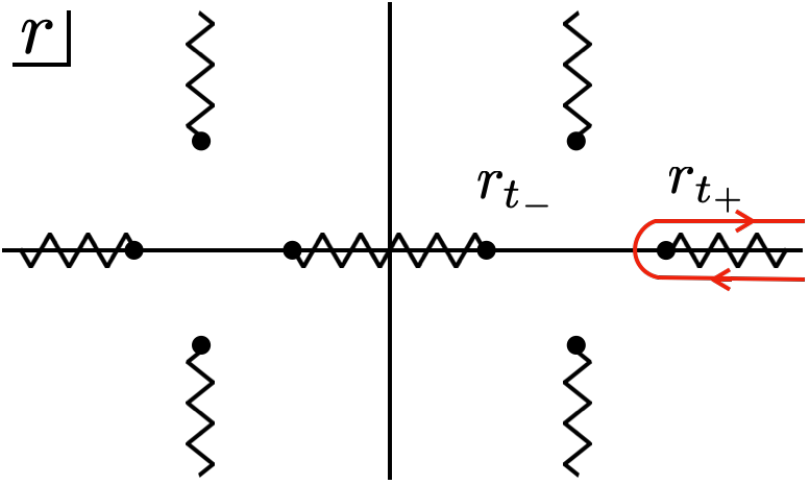}. \nn
\end{align}
where $r_{t_+}$ is the turning point outside the outer horizon and $r_{t_-}$ is the smaller turning point behind the inner horizon. The other branch points are the complex roots.

\begin{figure}[t]
 \includegraphics[scale=0.3, valign = c]{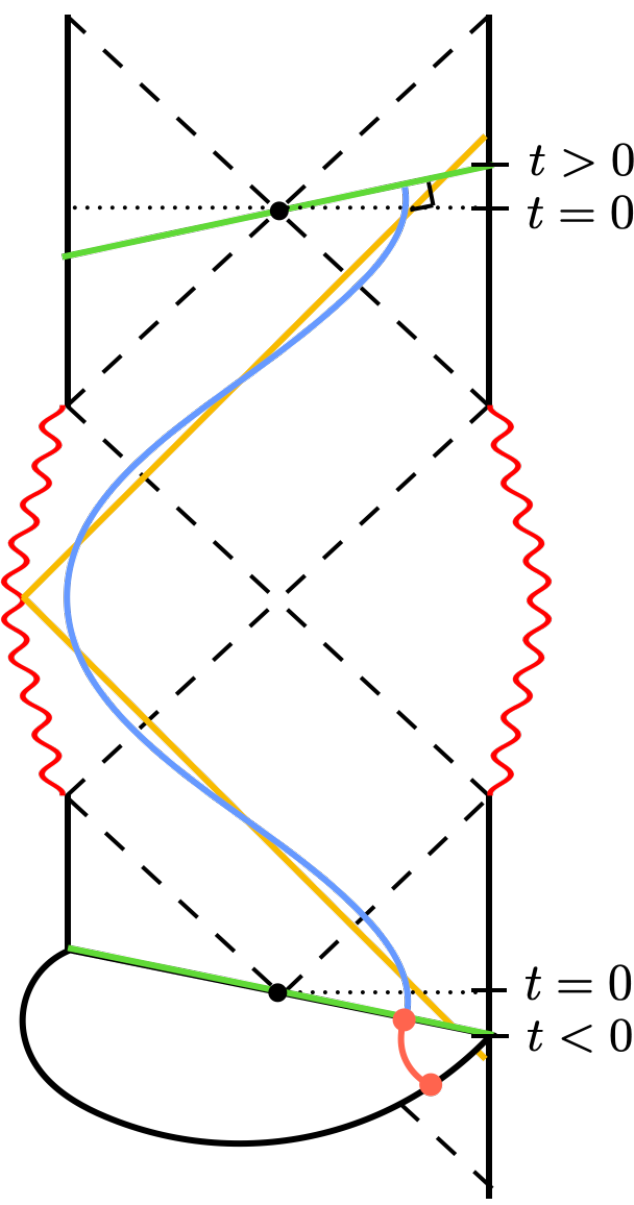}
\caption{\label{fig:RNShift}
{\footnotesize This shows the trajectory of geodesics that leave the Euclidean section after continuation. They generically emerge from times different from zero. }}
\end{figure}

A timelike excursion is generated by winding $r_{t_-}$ and $r_{t_+}$ around each other while avoiding the  other branch cuts. This is done by moving $E$ along a path in the complex $E$ plane around a branch point on the imaginary axis, as shown below on FIG.~\ref{fig:RNShift}(a).
In the large $r_+$ and small $L$ limit, the branch point is  $E^* = i r_+^4/2r_-$. 

Next, we look at coordinate shifts accompanying the timelike excursion. The Euclidean time coordinate is given by
\begin{align}
    \tau = E\int {dr \over f(r) \sqrt{ f(r) \left(1 -L^2 /r^2 \right) -E^2 }}. \label{tRN}
\end{align}
The integrand has poles at $\pm r_+$, $\pm r_-$, and $\pm i r_c$ where $r_c  \equiv  \sqrt{1 + r_-^2 + r_+^2}$, and eight branch points from the roots of the function in the square root. 
The contour after the continuation looks like
\begin{align}
  \includegraphics[scale=0.33, valign = c]{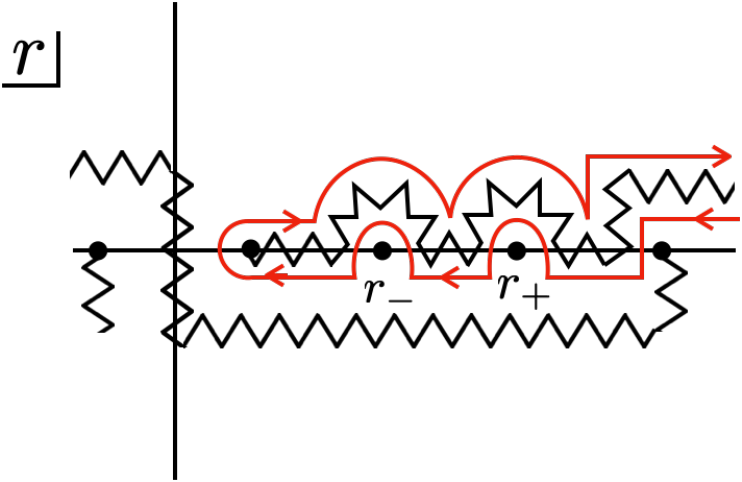}. \nn
\end{align}
We see that $\tau$ picks up two types of contributions. A pair of real contributions from the poles at the two horizons that evaluate to 
\begin{align}
 { \pi r_+^3 \over (r_+^2 - r_-^2)(1 + r_-^2 + 2 r_+^2)} -{  \pi r_-^3 \over (r_+^2 - r_-^2)(1 + 2 r_-^2 + r_+^2)}   \label{eq:RNHorizontshift}
\end{align}
which corresponds to $(\beta_+ - \beta_- )/2$, the inverse temperatures of the outer and inner horizons. These contributions can be thought of as penalities for crossing the two horizons; $2 \times \beta_+/4$ and $2 \times \beta_-/4$ for  the outer and inner horizons respectively. Note that they are independent of the boundary endpoints.

The second contribution is imaginary and comes from the segment parallel to the real axis. This gives a Lorentzian time shift that we call $t_s(E)$. The interpretation of this imaginary shift is that the geodesics start and end at different external times in the two universes before extending the Euclidean section; geodesics with an inner turning point at $t = 0$ start at a negative value of $t$ in the past universe and an equal but opposite sign time in the future universe; see FIG.~\ref{fig:RNShift}. Note that this effect was absent in BTZ.

We can obtain an analytic expression for this shift in the null limit. We calculate the coordinate time a null geodesic starting at the singularity hits the asymptotic boundary. For the case of $L = 0$, this is given by the integral
\begin{align}
	t = \int_0^r {dr \over f(r)}.
\end{align}
 This is the coordinate time shift for a null geodesic without angular momentum. The above can also be obtained from the strict $E \to \infty$ limit of Eq.~\eqref{tRN}. The trick to evaluating this is to consider the longer integration contour computing $2t$
 \begin{align}
    \includegraphics[scale=0.33, valign = c]{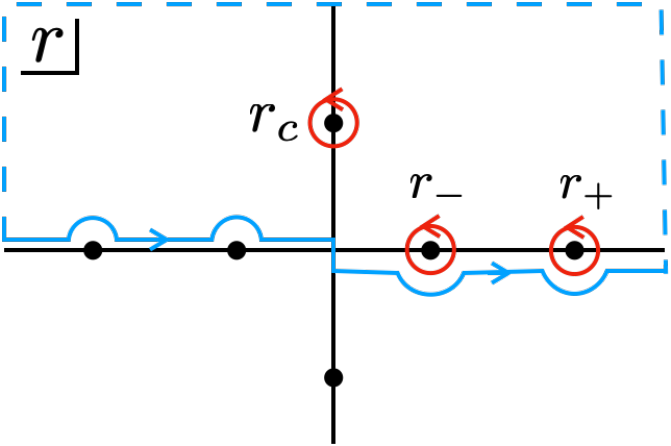}
\end{align}
and close the contour in the upper half plane, picking up the residues of the poles from the three zeros of $f(r)$, at $r =\{ r_-, r_+, ir_c \}$. The first two contributions reproduce \eqref{eq:RNHorizontshift} and the final one evaluates to
 \begin{align}
     {i\pi (1 + r_-^2 + r_+^2)^{3/2} \over (1 + 2 r_-^2 + r_+^2)(1 + r_-^2 + 2 r_+^2)},
 \end{align}
corresponding to half the inverse temperature $\beta_{c}$ of the imaginary horizon radius $ir_c$.
The null geodesic reaches the future asymptotic boundary at a positive time, implying that the singularity bends outwards! Therefore, the imaginary shift for timelike geodesics has the same interpretation; they become orthogonal to the $t = t_s(E)$ slice in the exterior, as shown in FIG.~\ref{fig:RNShift}, after which they extend into the Euclidean section.

The change in the angular coordinate is given by
\begin{align}
     \varphi = L\int {dr \over r^2 \sqrt{ f(r) \left( 1 - L^2 f(r)/r^2\right) - E^2}}.
\end{align}
Since there are no poles in the integrand, the angle does not pick up real shifts. Furthermore, the imaginary part vanishes in the $L \to 0$ limit since the turning point is away from $r = 0$.

\begin{figure}
    \centering
    \includegraphics[width=\linewidth]{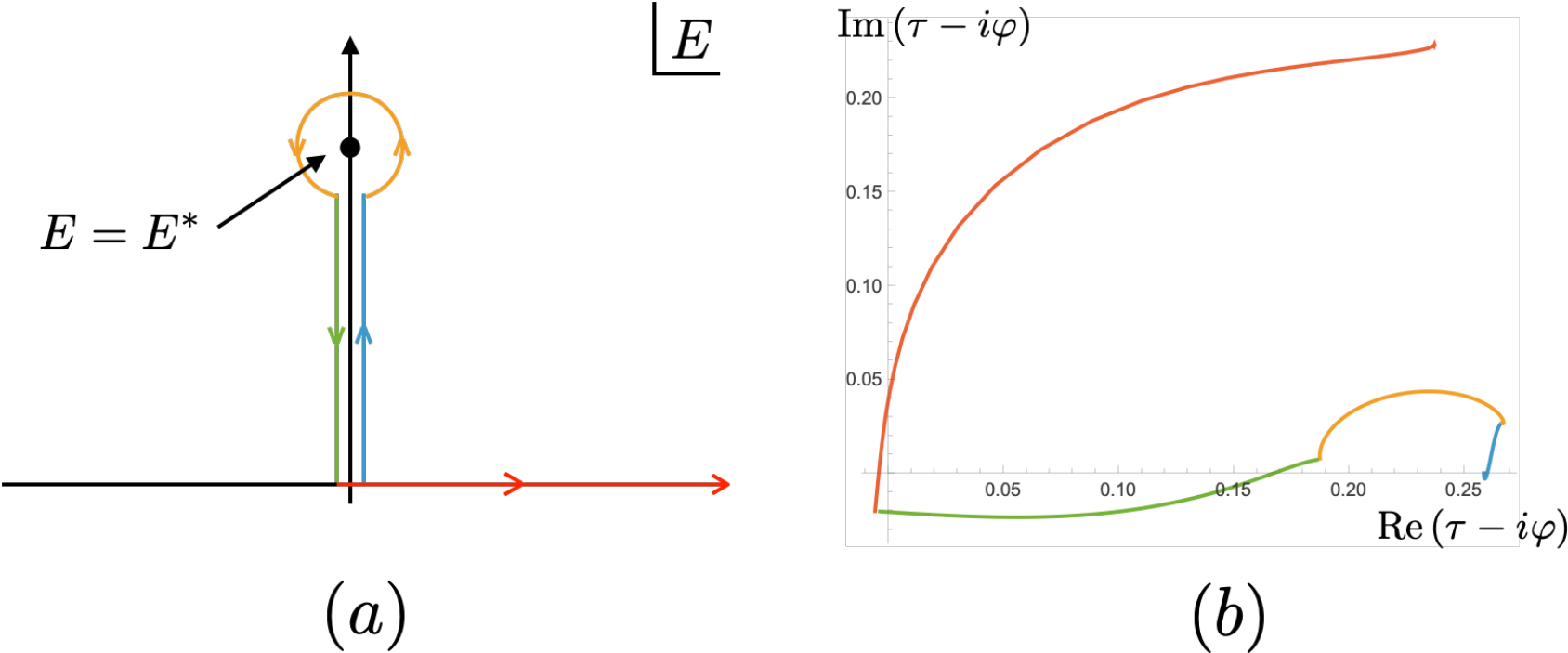}
    \caption{(a) The path of analytic continuation on the $E$ plane to generate the null geodesic that probes the singularity. We start with $E=0$ and $L=i\epsilon$ and wind around the branch point $E=E^*$ before sending $E\to \infty$. (b) The corresponding change in the $\tau$ coordinate. Different stages of analytic continuation are colored the same in the two figures. The path in the right pane winds around the branch point at $\tau-i\varphi=0$. As one sends $E\to\infty$ we have $\varphi\to0$ and $\tau\to\frac{1}{2}(\beta_+-\beta_-+i\beta_c)$. This plot is generated with $L=0.1i$ and $(r_+,r_-)=(\sqrt{10},1)$.}
    \label{fig:RNTimeContour}
\end{figure}

The upshot for the case of vanishing angular momentum $L = 0$, is that the analytic continuation in the coordinates is purely a shift of the time coordinate:
\begin{align}
    \tau \to \tau + \beta_+ -\beta_-   + i t_s(\tau), \label{eq:RNtimeshift}
\end{align}
where we invert the $E(\tau)$ to write $t_s$ as a function of $\tau$. The path along which this continuation has to be implemented is determined by the path of the geodesic energy continuation that swaps the turning points. In FIG.~\ref{fig:RNTimeContour} we show the case starting with a zero energy geodesic (with very small $L$) where $\tau = \beta_+/2$ and then taking the energy to infinity after rotating once around a branch point on the imaginary axis (FIG.~\ref{fig:RNTimeContour}(a)). Instead of getting the usual coincident singularity where $\tau \to i \varphi$, $\tau$ winds around $i \varphi$ and tends to $\tau = (\beta_+ - \beta_- + i \beta_c)/2$, as predicted. We expect that additional poles correspond to null geodesics that bounce between the boundary and the singularity multiple times as they traverse multiple universes.

  The WKB phase in momentum space can be obtained directly from the geodesic analysis above, and we expect the position space analysis to work analogously to BTZ. However, unlike BTZ, we do not have access to a closed form expression for the correlators at this point. Nevertheless, we can already see from the geodesic analysis above that the semi-classical momentum space correlator has branch cuts on the $E$ plane, spoiling periodicity. Similar to BTZ, we expect those to be resolved by a non-perturbative sum over images around the branch point, which corresponds to the timelike geodesic contributions seen above. We leave a deeper analysis of this to future work.

\subsection{{AdS Kerr black hole}}
As another example of our prescription, we consider the higher dimensional analog of AdS-Kerr black holes (also known as Myers-Perry black holes \cite{Myers:1986un}). Rotating black holes in $d$ spacetime dimensions have $d-3$ independent angular momenta. Again we focus on $d=5$ here, and for simplicity we assume the two angular momenta are equal.

The radial geodesic equation reads \cite{Delsate:2015ina,Grunau:2017uzf}
\begin{equation}
     \rho^6\dot{\rho}^2 = R(\rho),
\end{equation}
where $\rho\in(0,\infty)$ labels the radial direction. The radial turning points satisfy the condition $R(\rho)=0$.
$R$ is a quartic polynomial of $\rho^2$ whose detailed form along with the equations of motion of other coordinate components are given in Appendix~\ref{app:Kerr}.

Geodesics in this spacetime are specified by the particle's rest mass $m^2$, its energy $E$, the two angular momenta $L_\phi$ and $L_\psi$, plus an additional constant of motion $K$ known as Carter's constant \cite{Carter:1968rr}.
We show in Appendix~\ref{app:Kerr} that for the following choice of charges
\begin{align}
\begin{split}
    \quad &L_\phi=L\sin^2\chi, \quad L_\psi=L\cos^2\chi,\\
    \quad &K=-L^2-a^2(E^2-L^2)+a^2(m^2+k),
\end{split}
\end{align}
the particle has two turning points $\rho_{t_\pm}$ given by
\begin{equation}
    \rho_{t_+} \approx \sqrt{(1-a^2)(E^2+L^2)}, \quad \rho_{t_-} \approx 2a^4Mk
\end{equation}
in the limit where $E\gg 1$ and $k\ll 1$.
Here $a^2<1$ is related to the angular momenta of the black hole and $\chi$ labels the azimuthal angle of the worldline. Also note that $L^2<0$ since we are working with the Euclidean metric.

The geodesic length is given by
\begin{align}
    D = \int \frac{\rho^3}{\sqrt{R(\rho)}}d \rho 
\end{align}
The analytic structure of the integrand closely resembles that of the charged black hole considered above. 
Similarly, we expect that there exists some analytic continuation of the energy of $(E,L,k)$ that exchanges the two physical turning points $\rho_{t_\pm}$, under which the contour integral for $D$ picks up an imaginary contribution that probes the Lorentzian part of the spacetime. 
The detailed form of the continuation can in principle be constructed from the action of monodromy group of the polynomial $R(\rho)$ and we leave it to future work.

\begin{figure}[h]
    \centering
    \includegraphics[width=.85\linewidth]{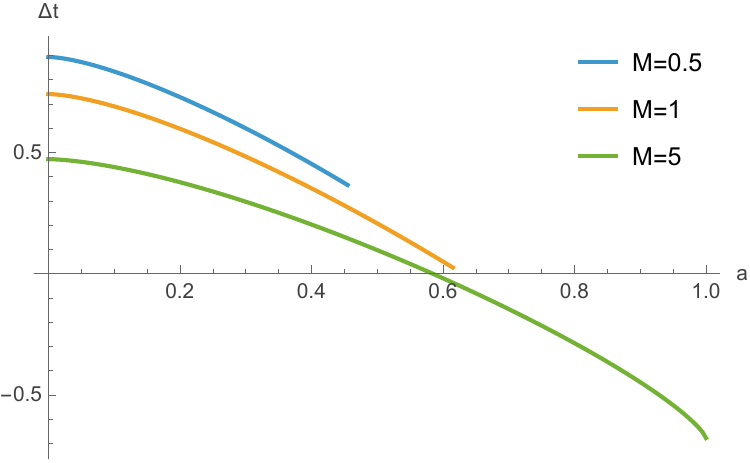}
    \caption{The elapsed coordinate time $\Delta t$ for the $L=0$ null geodesic considered in this section, plotted against different black hole mass $M$ and angular momenta parameter $a$. For each $M$, there is a different  $a^{\max}_M$ in the sense that the black hole starts to develop naked singularity for $a>a^{\max}_M$. The sign of $\Delta t$ determines the ``shape'' of the singularity in the extended Penrose diagram. When $\Delta t>0$, the singularity bends outward (see e.g. FIG.~\ref{RNIntro}). The singularity bends inwards for $\Delta t<0$.}
    \label{fig:dtKerr}
\end{figure}

For now, let us assume that the appropriate analytic continuation has been carried out. Since we work with $E\gg 1$, the corresponding timelike geodesic in the Lorentzian portion of the spacetime can be thought of as approximating a null ray that bounces off the curvature singularity. The change of the coordinate time along the Lorentzian portion can be calculated by
\begin{equation}
\label{eq:tKerr}
    \Delta t = \int^{\rho_{t_+}}_{\rho_{t_-}} \frac{\rho T(\rho)}{\sqrt{R(\rho)}}d\rho.
\end{equation}
The full form of $T(\rho)$ can be found in Appendix~\ref{app:Kerr}. The integral  can be evaluated in closed form using Weierstrass Elliptic functions (see \cite{Grunau:2017uzf}). Here we simply evaluate Eq.~\eqref{eq:tKerr} numerically and present the result in FIG.~\ref{fig:dtKerr}.

\section{{Black holes without inner horizons}}

All the black holes considered above featured an inner horizon region that connects distinct timelike-separated asymptotic regions, allowing for two real positive turning points. Additionally, we found there can be imaginary turning points corresponding to complex geodesics. The analytic continuation we propose is able to exchange any pair of turning points, including a real with a complex root. In this section, we analyze these complex geodesics in black holes without inner horizons with a single real turning point.

\subsection{Non-rotating BTZ black hole}
The simplest example is the non-rotating BTZ black hole,  obtained from Eq.~\eqref{euclideanBTZ} by setting the inner horizon radius $r_{-}$ to zero. The geodesic length formula remains unchanged while the boundary coordinate separations simplify as
\begin{gather}
    D(E,L) = -{1 \over 2}\ln A_+A_-\bar{A}_+\bar{A}_-,\\
    \tau = {\beta \over 4 \pi i} \ln {A_+ \bar{A}_+ \over A_- \bar{A}_-}, \quad \varphi = {\beta \over 4 \pi } \ln {A_+ \bar{A}_- \over A_- \bar{A}_+}.
\end{gather}
In this limit, there is an imaginary turning point in addition to the real turning point outside of the horizon. Interchanging them by rotating $E + i(L \pm  r_+)$ around $0$ modifies the geodesic length integration contour to
\begin{equation}
  \includegraphics[scale=0.33, valign = c]{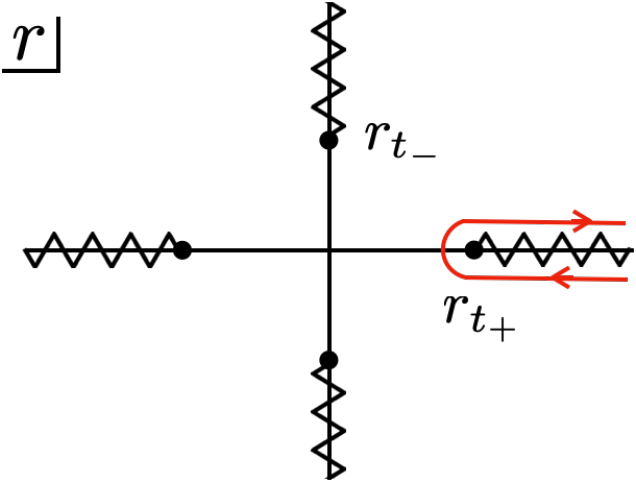} ~\to \includegraphics[scale=0.33, valign = c]{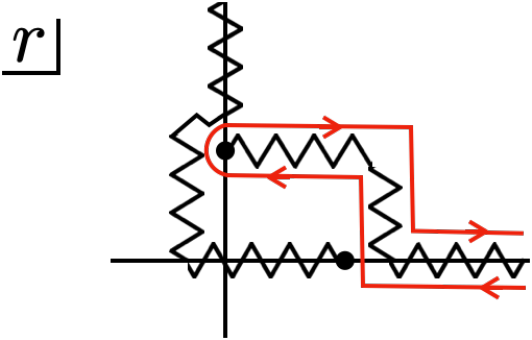},
\end{equation}
where $r_{t_-} = - i |r_{t_-}|$ is the negative branch of the imaginary turning point.

This continuation induces a position space continuation identical to rotating BTZ, taking $\tau \to \tau + \beta/2$ and $\varphi \to \varphi + i \beta/2$. However it is a complexified geodesic because the turning point is imaginary. Nevertheless, in the $E \to \infty$ limit, the turning point limits to the conical singularity at $r = 0$, at which the  geodesic becomes a real null geodesic. This null geodesic connects the right boundary in one BTZ universe to the left boundary in second BTZ universe glued along the future conical singularity, as shown here:
\begin{equation}
    \includegraphics[scale=.33, valign=c]{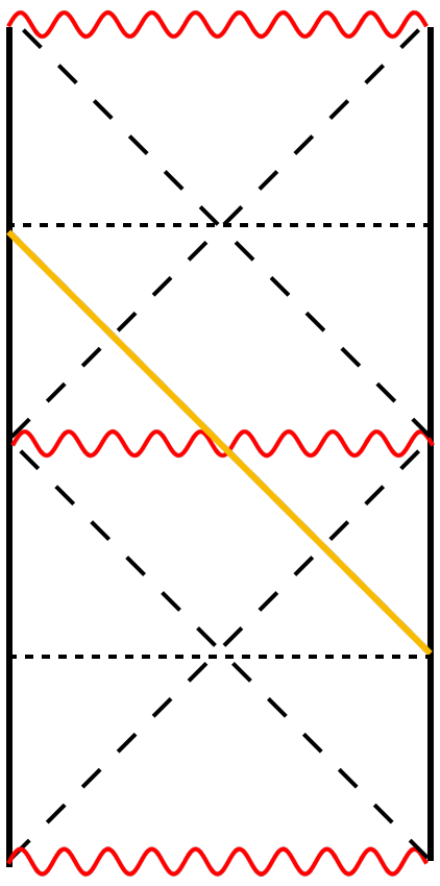}
\end{equation}

\subsection{AdS-Schwarzchild}
We also consider the Schwarzschild black hole in five dimensions, which can be obtained from Eq.~\eqref{EuclideanRN} by setting the charge $Q$ to zero. For simplicity, we focus on geodesics with $L = 0$ whose turning point equation is quadratic in $x \equiv r^2$ given by
\begin{equation}
    0= x^2 + (1- E^2 ) x - r_{0}^2 (1+ r_{0}^2),
\end{equation}
where $r_0$ is the event horizon radius. The equation has two real roots with opposite signs in $x$, indicating an imaginary turning radius in addition to the exterior turning radius. 

For the case of $L = 0$, the geodesic length and boundary time separation can be evaluated exactly \cite{Festuccia:2005pi}
\begin{gather}
    D(E) = -{1 \over 2} \ln \left({\cal{A}}_+  {\cal{A}}_- \bar{{\cal{A}}}_+ \bar{{\cal{A}}}_- \right), \\
    t(E) = {\beta \over 4 \pi} \ln {{\cal{A}}_+ \bar{{\cal{A}}}_- \over{\cal{A}}_- \bar{{\cal{A}}}_+} - {i\tilde{\beta} \over 4 \pi}\ln {{\cal{A}}_+ \bar{{\cal{A}}}_+ \over {\cal{A}}_- \bar{{\cal{A}}}_-} - {i \beta \over 2}
\end{gather}
where
\begin{align}
    {\cal{A}}_\pm = {1 \over 2} \pm {\tilde{\beta} + i \beta \over 4 \pi } E, \quad \bar{{\cal{A}}}_\pm = {1 \over 2} \pm {\tilde{\beta} - i \beta \over 4 \pi } E.
\end{align}
In our prescription, interchanging the real and imaginary turning points amounts to rotating one of the ${\cal{A}}$'s, say ${\cal{A}}_-$, around the origin. Going around once shifts $D \to D-i\pi$ and $t \to t + (i\beta + \tilde{\beta})/2$. The time shift is the geometric signature of the boundary singularity \cite{Fidkowski:2003nf}. The resulting geodesic is complex. However, just like in non-rotating BTZ, the complex turning point tends to the singularity in the large $E$ limit. As a confirmation, the boundary time separation between the operators is precisely the predicted shift $(i\beta + \tilde{\beta})/2$. Therefore, the bouncing geodesic is retrieved as a limit of complex geodesics.

\begin{figure}[t] 
    \centering
    \includegraphics[width=1\linewidth]{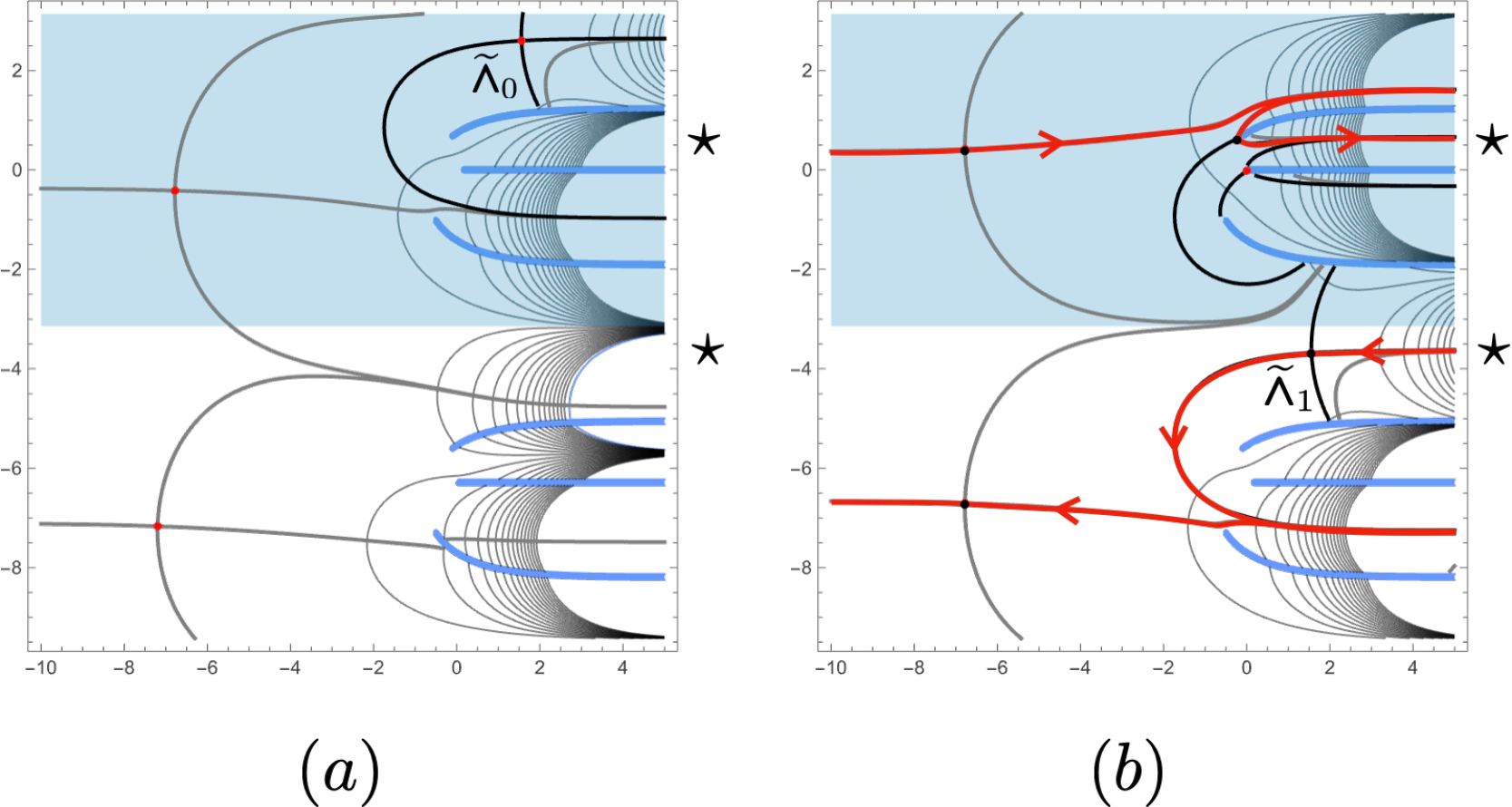}
    \caption{Paths of steepest descent for the Fourier transform of the AdS-Schwarzschild momentum space correlator. We work in the coordinate system where the branch cut associated to $\mathcal{A}_-$ is unwrapped. The principal sheet is shaded in blue. (a) The $n=0$ bouncing geodesic saddle $\widetilde{\pmb{\wedge}}_0$ is not on the steepest descent contour connecting the defining asymptotic regions. (b) The $n=1$ contribution from the image sum around ${\cal A}_- = 0$ gives a displaced bouncing geodesic saddle $\widetilde{\pmb{\wedge}}_1$ that lies on the principal steepest descent contour.}
    \label{fig:StokesSch}
\end{figure}

It was found in \cite{Fidkowski:2003nf,Festuccia:2005pi} that the bouncing singularity saddle does not lie on the integration contour. We reproduce this in FIG.~\ref{fig:StokesSch}(a) where the bouncing geodesic saddle $\widetilde{\pmb{\wedge}}_0$ \footnote{The symbol $\widetilde{\pmb{\wedge}}$ is chosen to resemble a bouncing geodesic reflected by the spacelike singularity.} is not captured by the correct steepest descent path. However, the structure of $D(E)$ and $t(E)$ is very reminiscent of BTZ, with branch cuts that spoil periodicity around ${\cal A}_- = 0$. We therefore expect that one has to sum over images around the branch point once the argument of ${\cal A}_-$ is large enough, which would pick up the bouncing geodesic. Indeed, we find that adding the $n = 1$ image also gives a bouncing geodesic saddle $\widetilde{\pmb{\wedge}}_1$, but one that is shifted relative to $\widetilde{\pmb{\wedge}}_0$ and lies on the correct steepest descent path as shown in FIG.~\ref{fig:StokesSch}(b). Recall that this comes from the $E \to \infty$ limit of complex geodesics which are analytic continuations of real exterior geodesics, and therefore should be solutions to the original WKB problem. It would be interesting to see these image contributions directly from solutions to the wave equation, along the lines discussed in \cite{Afkhami-Jeddi:2025wra}.
\section{{Outlook}}

In this paper we showed how to extract analytic continuations from the WKB limit of the boundary anchored correlators to probe past the inner horizon. In particular, we saw how to reproduce timelike contributions to the geodesic length from paths that connect two timelike related universes through the interior of rotating and charged black holes.

What makes these results potentially useful is their ability to probe the inner horizons of charged and rotating black holes. Beyond the strict classical picture with no perturbations, the inner horizon is expected to be classically unstable~\cite{SimpsonPenrose1973, McNamara1978,ChandrasekharHartle1982,PoissonIsrael1990,Ori1991} and outright absent due to quantum effects~\cite{Hiscock:1977wm,Birrell:1978th,Hollands:2019whz}. 

Classically, generic matter and gravitational perturbations lead to divergences near the inner horizon \cite{SimpsonPenrose1973, McNamara1978,ChandrasekharHartle1982,PoissonIsrael1990,Ori1991}. It would be interesting to see whether signatures of this instability can arise from probe corrections to the two-point function's analytic structure. An example of such a phenomenon observed previously led to the resolution of the forbidden singularities problem \cite{Fitzpatrick:2016ive} in two-point functions in a black hole microstate~\cite{Fitzpatrick:2016ive,Fitzpatrick:2016mjq,Faulkner:2017hll, Chen:2017yze, Yang:2018gdb}. We hope to return to this problem in the near future.

Beyond the classical picture, the quantum stress-energy tensor is expected to diverge near the inner horizon of charged and rotating black holes~\cite{Hiscock:1977wm,Birrell:1978th,Hollands:2019whz}. As a first step to a more complete treatment, one can incorporate the effects of treating matter quantum-mechanically by a systematic WKB expansion beyond the geodesic approximation. We do so for the five-dimensional AdS Reissner-Nordstrom black hole in Appendix~\ref{app:WKBbreakdown}, and find signatures of the breakdown of the WKB method near the singularity in the large energy limit. Accounting for quantum corrections for the geometry involves solving a difficult backreaction problem, although some progress has been made in certain models. A concrete example of this is a black hole in braneworld holography~\cite{Emparan:2002px,Emparan:2020znc}. The solution studied there is a quantum-corrected BTZ black hole, which are known to also admit bouncing geodesics \cite{Kolanowski:2023hvh}. In future work~\cite{future}, we analyze this problem from the perspective of the holographic duality on the braneworld.

\section*{{Acknowledgements}}
We thank Simon Caron-Huot, Adam Levine, Hong Liu, Juan Maldacena, Shoy Ouseph, Malcolm Perry, and Stephen Shenker for discussion on this topic. AA acknowledges support of this work from the J. Robert Oppenheimer Endowed Fund and Tamkeen under the NYU Abu Dhabi Research Award (ADHPG -- AD375). SAA acknowledges support from the James Arthur Graduate Associate fellowship. 

\bibliographystyle{ourbst}
\bibliography{post.bib}

@article{Fidkowski:2003nf,
    author = "Fidkowski, Lukasz and Hubeny, Veronika and Kleban, Matthew and Shenker, Stephen",
    title = "{The Black hole singularity in AdS / CFT}",
    eprint = "hep-th/0306170",
    archivePrefix = "arXiv",
    reportNumber = "SU-ITP-03-16",
    doi = "10.1088/1126-6708/2004/02/014",
    journal = "JHEP",
    volume = "02",
    pages = "014",
    year = "2004"
}

@article{Festuccia:2005pi,
    author = "Festuccia, Guido and Liu, Hong",
    title = "{Excursions beyond the horizon: Black hole singularities in Yang-Mills theories. I.}",
    eprint = "hep-th/0506202",
    archivePrefix = "arXiv",
    reportNumber = "MIT-CTP-3641",
    doi = "10.1088/1126-6708/2006/04/044",
    journal = "JHEP",
    volume = "04",
    pages = "044",
    year = "2006"
}

@article{Festuccia:2008zx,
    author = "Festuccia, Guido and Liu, Hong",
    title = "{A Bohr-Sommerfeld quantization formula for quasinormal frequencies of AdS black holes}",
    eprint = "0811.1033",
    archivePrefix = "arXiv",
    primaryClass = "gr-qc",
    reportNumber = "MIT-CTP-3995, SCIPP-08-11",
    doi = "10.1166/asl.2009.1029",
    journal = "Adv. Sci. Lett.",
    volume = "2",
    pages = "221--235",
    year = "2009"
}

@article{Ceplak:2024bja,
    author = "{\v{C}}eplak, Nejc and Liu, Hong and Parnachev, Andrei and Valach, Samuel",
    title = "{Black hole singularity from OPE}",
    eprint = "2404.17286",
    archivePrefix = "arXiv",
    primaryClass = "hep-th",
    doi = "10.1007/JHEP10(2024)105",
    journal = "JHEP",
    volume = "10",
    pages = "105",
    year = "2024"
}

@article{Dodelson:2023nnr,
    author = "Dodelson, Matthew and Iossa, Cristoforo and Karlsson, Robin and Lupsasca, Alexandru and Zhiboedov, Alexander",
    title = "{Black hole bulk-cone singularities}",
    eprint = "2310.15236",
    archivePrefix = "arXiv",
    primaryClass = "hep-th",
    reportNumber = "CERN-TH-2023-192",
    doi = "10.1007/JHEP07(2024)046",
    journal = "JHEP",
    volume = "07",
    pages = "046",
    year = "2024"
}

@article{Dodelson:2023vrw,
    author = "Dodelson, Matthew and Iossa, Cristoforo and Karlsson, Robin and Zhiboedov, Alexander",
    title = "{A thermal product formula}",
    eprint = "2304.12339",
    archivePrefix = "arXiv",
    primaryClass = "hep-th",
    reportNumber = "CERN-TH-2023-062",
    doi = "10.1007/JHEP01(2024)036",
    journal = "JHEP",
    volume = "01",
    pages = "036",
    year = "2024"
}

@article{Afkhami-Jeddi:2025wra,
    author = "Afkhami-Jeddi, Nima and Caron-Huot, Simon and Chakravarty, Joydeep and Maloney, Alexander",
    title = "{Imprint of the black hole singularity on thermal two-point functions}",
    eprint = "2510.21673",
    archivePrefix = "arXiv",
    primaryClass = "hep-th",
    month = "10",
    year = "2025"
}

@article{Dodelson:2024atp,
    author = "Dodelson, Matthew",
    title = "{Ringdown in the SYK model}",
    eprint = "2408.05790",
    archivePrefix = "arXiv",
    primaryClass = "hep-th",
    reportNumber = "CERN-TH-2024-135",
    doi = "10.21468/SciPostPhys.19.3.081",
    journal = "SciPost Phys.",
    volume = "19",
    pages = "081",
    year = "2025"
}

@article{Balasubramanian:2004zu,
    author = "Balasubramanian, Vijay and Levi, Thomas S.",
    title = "{Beyond the veil: Inner horizon instability and holography}",
    eprint = "hep-th/0405048",
    archivePrefix = "arXiv",
    reportNumber = "UPR-T-1079",
    doi = "10.1103/PhysRevD.70.106005",
    journal = "Phys. Rev. D",
    volume = "70",
    pages = "106005",
    year = "2004"
}

@article{Balasubramanian:2019qwk,
    author = "Balasubramanian, Vijay and Kar, Arjun and S{\'a}rosi, G{\'a}bor",
    title = "{Holographic Probes of Inner Horizons}",
    eprint = "1911.12413",
    archivePrefix = "arXiv",
    primaryClass = "hep-th",
    reportNumber = "CERN-TH-2019-207",
    doi = "10.1007/JHEP06(2020)054",
    journal = "JHEP",
    volume = "06",
    pages = "054",
    year = "2020"
}

@article{Kravchuk:2018htv,
    author = "Kravchuk, Petr and Simmons-Duffin, David",
    title = "{Light-ray operators in conformal field theory}",
    eprint = "1805.00098",
    archivePrefix = "arXiv",
    primaryClass = "hep-th",
    reportNumber = "CALT-TH 2018-018",
    doi = "10.1007/JHEP11(2018)102",
    journal = "JHEP",
    volume = "11",
    pages = "102",
    year = "2018"
}

@article{Faulkner:2017hll,
    author = "Faulkner, Thomas and Wang, Huajia",
    title = "{Probing beyond ETH at large $c$}",
    eprint = "1712.03464",
    archivePrefix = "arXiv",
    primaryClass = "hep-th",
    doi = "10.1007/JHEP06(2018)123",
    journal = "JHEP",
    volume = "06",
    pages = "123",
    year = "2018"
}

@article{Emparan:2020znc,
    author = "Emparan, Roberto and Frassino, Antonia Micol and Way, Benson",
    title = "{Quantum BTZ black hole}",
    eprint = "2007.15999",
    archivePrefix = "arXiv",
    primaryClass = "hep-th",
    doi = "10.1007/JHEP11(2020)137",
    journal = "JHEP",
    volume = "11",
    pages = "137",
    year = "2020"
}

@article{Kolanowski:2023hvh,
    author = "Kolanowski, Maciej and Toma{\v{s}}evi{\'c}, Marija",
    title = "{Singularities in 2D and 3D quantum black holes}",
    eprint = "2310.06014",
    archivePrefix = "arXiv",
    primaryClass = "hep-th",
    reportNumber = "CPHT-RR064.102023",
    doi = "10.1007/JHEP12(2023)102",
    journal = "JHEP",
    volume = "12",
    pages = "102",
    year = "2023"
}

@article{Hollands:2019whz,
    author = "Hollands, Stefan and Wald, Robert M. and Zahn, Jochen",
    title = {{Quantum instability of the Cauchy horizon in Reissner{\textendash}Nordstr{\"o}m{\textendash}deSitter spacetime}},
    eprint = "1912.06047",
    archivePrefix = "arXiv",
    primaryClass = "gr-qc",
    doi = "10.1088/1361-6382/ab8052",
    journal = "Class. Quant. Grav.",
    volume = "37",
    number = "11",
    pages = "115009",
    year = "2020"
}

@article{Grunau:2017uzf,
    author = "Grunau, Saskia and Neumann, Hendrik and Reimers, Stephan",
    title = "{Geodesic motion in the five-dimensional Myers-Perry-AdS spacetime}",
    eprint = "1711.02933",
    archivePrefix = "arXiv",
    primaryClass = "gr-qc",
    doi = "10.1103/PhysRevD.97.044011",
    journal = "Phys. Rev. D",
    volume = "97",
    number = "4",
    pages = "044011",
    year = "2018"
}

@article{Carter:1968rr,
    author = "Carter, Brandon",
    title = "{Global structure of the Kerr family of gravitational fields}",
    doi = "10.1103/PhysRev.174.1559",
    journal = "Phys. Rev.",
    volume = "174",
    pages = "1559--1571",
    year = "1968"
}

@article{Myers:1986un,
    author = "Myers, Robert C. and Perry, M. J.",
    title = "{Black Holes in Higher Dimensional Space-Times}",
    reportNumber = "PRINT-86-0067 (PRINCETON)",
    doi = "10.1016/0003-4916(86)90186-7",
    journal = "Annals Phys.",
    volume = "172",
    pages = "304",
    year = "1986"
}

@article{Hawking:1998kw,
    author = "Hawking, S. W. and Hunter, C. J. and Taylor, Marika",
    title = "{Rotation and the AdS / CFT correspondence}",
    eprint = "hep-th/9811056",
    archivePrefix = "arXiv",
    doi = "10.1103/PhysRevD.59.064005",
    journal = "Phys. Rev. D",
    volume = "59",
    pages = "064005",
    year = "1999"
}

@article{Delsate:2015ina,
    author = "Delsate, T{\'e}rence and Rocha, Jorge V. and Santarelli, Raphael",
    title = "{Geodesic motion in equal angular momenta Myers-Perry-AdS spacetimes}",
    eprint = "1507.03602",
    archivePrefix = "arXiv",
    primaryClass = "gr-qc",
    doi = "10.1103/PhysRevD.92.084028",
    journal = "Phys. Rev. D",
    volume = "92",
    number = "8",
    pages = "084028",
    year = "2015"
}

@article{Gonzalez:2023jhx,
    author = "Gonz{\'a}lez, P. A. and Olivares, Marco and V{\'a}squez, Yerko and Villanueva, J. R.",
    title = {{Time like geodesics for five-dimensional Schwarzschild and Reissner{\textendash}Nordstr{\"o}m anti-de Sitter black holes}},
    eprint = "2308.01498",
    archivePrefix = "arXiv",
    primaryClass = "gr-qc",
    doi = "10.1140/epjc/s10052-023-12018-4",
    journal = "Eur. Phys. J. C",
    volume = "83",
    number = "9",
    pages = "853",
    year = "2023"
}

@article{Gonzalez:2020zfd,
    author = "Gonz{\'a}lez, P. A. and Olivares, Marco and V{\'a}squez, Yerko and Villanueva, J. R.",
    title = "{Null geodesics in five-dimensional Reissner-Nordstr$\ddot{o}$m anti-de Sitter black hole}",
    eprint = "2010.01442",
    archivePrefix = "arXiv",
    primaryClass = "gr-qc",
    doi = "10.1140/epjc/s10052-021-09024-9",
    journal = "Eur. Phys. J. C",
    volume = "81",
    number = "3",
    pages = "236",
    year = "2021"
}

@article{Cruz:1994ir,
    author = "Cruz, Norman and Martinez, Cristian and Pena, Leda",
    title = "{Geodesic structure of the (2+1) black hole}",
    eprint = "gr-qc/9401025",
    archivePrefix = "arXiv",
    doi = "10.1088/0264-9381/11/11/014",
    journal = "Class. Quant. Grav.",
    volume = "11",
    pages = "2731--2740",
    year = "1994"
}

@article{Banados:1992wn,
    author = "Banados, Maximo and Teitelboim, Claudio and Zanelli, Jorge",
    title = "{The Black hole in three-dimensional space-time}",
    eprint = "hep-th/9204099",
    archivePrefix = "arXiv",
    reportNumber = "PRINT-92-0151 (CHILE), IASSNS-HEP-92-29",
    doi = "10.1103/PhysRevLett.69.1849",
    journal = "Phys. Rev. Lett.",
    volume = "69",
    pages = "1849--1851",
    year = "1992"
}

@article{Ceplak:2025dds,
    author = "{\v{C}}eplak, Nejc and Liu, Hong and Parnachev, Andrei and Valach, Samuel",
    title = "{Fooling the Censor: Going beyond inner horizons with the OPE}",
    eprint = "2511.09638",
    archivePrefix = "arXiv",
    primaryClass = "hep-th",
    month = "11",
    year = "2025"
}

@article{Dodelson:2025jff,
    author = "Dodelson, Matthew and Iossa, Cristoforo and Karlsson, Robin",
    title = "{Bouncing off a stringy singularity}",
    eprint = "2511.09616",
    archivePrefix = "arXiv",
    primaryClass = "hep-th",
    month = "11",
    year = "2025"
}

@article{Witten:1998qj,
    author = "Witten, Edward",
    title = "{Anti de Sitter space and holography}",
    eprint = "hep-th/9802150",
    archivePrefix = "arXiv",
    reportNumber = "IASSNS-HEP-98-15",
    doi = "10.4310/ATMP.1998.v2.n2.a2",
    journal = "Adv. Theor. Math. Phys.",
    volume = "2",
    pages = "253--291",
    year = "1998"
}

@article{Gubser:1998bc,
    author = "Gubser, S. S. and Klebanov, Igor R. and Polyakov, Alexander M.",
    title = "{Gauge theory correlators from noncritical string theory}",
    eprint = "hep-th/9802109",
    archivePrefix = "arXiv",
    reportNumber = "PUPT-1767",
    doi = "10.1016/S0370-2693(98)00377-3",
    journal = "Phys. Lett. B",
    volume = "428",
    pages = "105--114",
    year = "1998"
}

@article{Hemming:2002kd,
    author = "Hemming, Samuli and Keski-Vakkuri, Esko and Kraus, Per",
    title = "{Strings in the extended BTZ space-time}",
    eprint = "hep-th/0208003",
    archivePrefix = "arXiv",
    reportNumber = "HIP-2002-30-TH, UCLA-02-TEP-21",
    doi = "10.1088/1126-6708/2002/10/006",
    journal = "JHEP",
    volume = "10",
    pages = "006",
    year = "2002"
}

@article{Kraus:2002iv,
    author = "Kraus, Per and Ooguri, Hirosi and Shenker, Stephen",
    title = "{Inside the horizon with AdS / CFT}",
    eprint = "hep-th/0212277",
    archivePrefix = "arXiv",
    reportNumber = "UCLA-02-TEP-41, CALT-68-2421, SU-ITP-02-45",
    doi = "10.1103/PhysRevD.67.124022",
    journal = "Phys. Rev. D",
    volume = "67",
    pages = "124022",
    year = "2003"
}

@article{Hartman:2013qma,
    author = "Hartman, Thomas and Maldacena, Juan",
    title = "{Time Evolution of Entanglement Entropy from Black Hole Interiors}",
    eprint = "1303.1080",
    archivePrefix = "arXiv",
    primaryClass = "hep-th",
    doi = "10.1007/JHEP05(2013)014",
    journal = "JHEP",
    volume = "05",
    pages = "014",
    year = "2013"
}

@article{Liu:2002kb,
    author = "Liu, Hong and Moore, Gregory W. and Seiberg, Nathan",
    title = "{Strings in time dependent orbifolds}",
    eprint = "hep-th/0206182",
    archivePrefix = "arXiv",
    reportNumber = "RUNHETC-2002-19, NI-02014-MTH",
    doi = "10.1088/1126-6708/2002/10/031",
    journal = "JHEP",
    volume = "10",
    pages = "031",
    year = "2002"
}

@article{Harlow:2011ny,
    author = "Harlow, Daniel and Maltz, Jonathan and Witten, Edward",
    title = "{Analytic Continuation of Liouville Theory}",
    eprint = "1108.4417",
    archivePrefix = "arXiv",
    primaryClass = "hep-th",
    reportNumber = "SU-ITP-11-42",
    doi = "10.1007/JHEP12(2011)071",
    journal = "JHEP",
    volume = "12",
    pages = "071",
    year = "2011"
}

@misc{future,
    author = "Ali Ahmad, Shadi and Almheiri, Ahmed and Lin,Simon and Toma{\v{s}}evi{\'c}, Marija",
    title = "{work in progress}",
}

@article{Emparan:2002px,
    author = "Emparan, Roberto and Fabbri, Alessandro and Kaloper, Nemanja",
    title = "{Quantum black holes as holograms in AdS brane worlds}",
    eprint = "hep-th/0206155",
    archivePrefix = "arXiv",
    reportNumber = "SU-ITP-02-23, CERN-TH-2002-131",
    doi = "10.1088/1126-6708/2002/08/043",
    journal = "JHEP",
    volume = "08",
    pages = "043",
    year = "2002"
}

@article{SimpsonPenrose1973,
  author  = {Simpson, M. and Penrose, R.},
  title   = {Internal instability in a {R}eissner--{N}ordstr{\"o}m black hole},
  journal = {International Journal of Theoretical Physics},
  volume  = {7},
  pages   = {183--197},
  year    = {1973},
  doi     = {10.1007/BF00792069}
}

@article{McNamara1978,
  author  = {McNamara, J. M.},
  title   = {Instability of black hole inner horizons},
  journal = {Proceedings of the Royal Society of London. Series A, Mathematical and Physical Sciences},
  volume  = {358},
  pages   = {499--517},
  year    = {1978},
  doi     = {10.1098/rspa.1978.0024}
}

@article{ChandrasekharHartle1982,
  author  = {Chandrasekhar, S. and Hartle, J. B.},
  title   = {On crossing the {C}auchy horizon of a {R}eissner--{N}ordstr{\"o}m black-hole},
  journal = {Proceedings of the Royal Society of London. Series A, Mathematical and Physical Sciences},
  volume  = {384},
  pages   = {301--315},
  year    = {1982},
  doi     = {10.1098/rspa.1982.0160}
}

@article{PoissonIsrael1990,
  author  = {Poisson, Eric and Israel, Werner},
  title   = {Internal structure of black holes},
  journal = {Physical Review D},
  volume  = {41},
  number  = {6},
  pages   = {1796--1809},
  year    = {1990},
  doi     = {10.1103/PhysRevD.41.1796}
}

@article{Ori1991,
  author  = {Ori, Amos},
  title   = {Inner structure of a charged black hole: An exact mass-inflation solution},
  journal = {Physical Review Letters},
  volume  = {67},
  number  = {7},
  pages   = {789--792},
  year    = {1991},
  doi     = {10.1103/PhysRevLett.67.789}
}

@article{Hiscock:1977wm,
    author = "Hiscock, W. A.",
    title = "{Stress Energy Tensor for a Two-Dimensional Evaporating Black Hole}",
    doi = "10.1103/PhysRevD.16.2673",
    journal = "Phys. Rev. D",
    volume = "16",
    pages = "2673--2674",
    year = "1977"
}

@article{Birrell:1978th,
    author = "Birrell, N. D. and Davies, P. C. W.",
    title = "{On falling through a black hole into another universe}",
    doi = "10.1038/272035a0",
    journal = "Nature",
    volume = "272",
    pages = "35",
    year = "1978"
}

@article{Fitzpatrick:2016mjq,
    author = "Fitzpatrick, A. Liam and Kaplan, Jared",
    title = "{On the Late-Time Behavior of Virasoro Blocks and a Classification of Semiclassical Saddles}",
    eprint = "1609.07153",
    archivePrefix = "arXiv",
    primaryClass = "hep-th",
    doi = "10.1007/JHEP04(2017)072",
    journal = "JHEP",
    volume = "04",
    pages = "072",
    year = "2017"
}

@article{Fitzpatrick:2016ive,
    author = "Fitzpatrick, A. Liam and Kaplan, Jared and Li, Daliang and Wang, Junpu",
    title = "{On information loss in AdS$_{3}$/CFT$_{2}$}",
    eprint = "1603.08925",
    archivePrefix = "arXiv",
    primaryClass = "hep-th",
    doi = "10.1007/JHEP05(2016)109",
    journal = "JHEP",
    volume = "05",
    pages = "109",
    year = "2016"
}

@article{Chen:2017yze,
    author = "Chen, Hongbin and Hussong, Charles and Kaplan, Jared and Li, Daliang",
    title = "{A Numerical Approach to Virasoro Blocks and the Information Paradox}",
    eprint = "1703.09727",
    archivePrefix = "arXiv",
    primaryClass = "hep-th",
    doi = "10.1007/JHEP09(2017)102",
    journal = "JHEP",
    volume = "09",
    pages = "102",
    year = "2017"
}

@article{Yang:2018gdb,
    author = "Yang, Zhenbin",
    title = "{The Quantum Gravity Dynamics of Near Extremal Black Holes}",
    eprint = "1809.08647",
    archivePrefix = "arXiv",
    primaryClass = "hep-th",
    doi = "10.1007/JHEP05(2019)205",
    journal = "JHEP",
    volume = "05",
    pages = "205",
    year = "2019"
}

\onecolumngrid
\appendix
\section{WKB approximation of scalar propagation in AdS black holes} \label{app: wkb}

Consider the following $D= d+1$-dimensional spacetime
\begin{equation}
    ds^2 = -f(r) dt^2 + f(r)^{-1}dr^2 + r^2 d\Omega_{d-1}^2,
\end{equation}
where $d\Omega_{d-1}^2$ is the line element on $S^{d-1}$. The d'Alembertian on this background is
\begin{align}
    \Box \phi &= \frac{1}{\sqrt{-g}} \partial_{\mu} \left[\sqrt{-g} g^{\mu \nu} \partial_{\nu} \phi \right], \\
    &= - f^{-1} \partial{t}^2 \phi + r^{1-d} \partial_{r} \left( r^{d-1} f \partial_{r} \phi\right) + \frac{1}{r^2} \Delta_{S^{d-1}} \phi,
\end{align}
where $\Delta_{S^{d-1}} $ is the Laplacian on the sphere. 

Let $Y_{\ell}(\Omega)$ be the spherical harmonic satisfying
\begin{equation}
    \Delta_{S^{d-1}}Y_{\ell } = - \ell ( \ell + d-2) Y_{\ell}, 
\end{equation}
and define the parameter
\begin{equation}
    \nu = \sqrt{ (\frac{d}{2})^2+ m^2}
\end{equation}
with $m$ being the mass of the neutral scalar. Because of the static and spherically symmetry nature of this metric, a convenient ansatz is
\begin{equation}
    \phi = e^{- i \omega t} Y_{\ell}(\Omega) \rho(r). 
\end{equation}
The Klein-Gordon equation becomes
\begin{equation}
r^{1-d} \partial_{r} \left( r^{d-1} f \partial_{r} \rho\right) + \left[ \frac{\omega^2}{f} - \frac{\ell( \ell + d -2)}{r^2} - m^2\right] \rho =0.
\end{equation}
Focusing on the first term, we may define the tortoise coordinate $\frac{d z}{dr}= f^{-1}$, so that it becomes
\begin{equation}
    r^{1-d} f^{-1}\partial_{z} \left( r^{d-1} \partial_{r_{*}} \rho\right)= f^{-1} \left[ \partial_{z}^2 \rho + (d-1) \frac{f}{r} \partial_{z}\rho  \right]. 
\end{equation}
Then, letting $\rho = r^{\frac{d-1}{2}} \psi$ cancels the first derivative term above and the full radial equation is recast as a Schrodinger equation
\begin{equation}
\partial_{z}^2 \psi + \left[\omega^2 - V_{\ell}(r) \right] \psi =0,
\end{equation}
where the potential is
\begin{equation}
    V_{\ell}(r) = f \left[m^2 + \frac{\ell ( \ell + d -2)}{r^2} + \frac{d-1}{2r } f' +\frac{(d-1)(d-3)}{4} \frac{f}{r^2 } \right].
\end{equation}
The potential entering the Schrodinger equation for AdS Reissner-Nordstrom in five dimensions is then given by
\begin{equation}
    V_{\ell}(r)= f(r) \left[\frac{(\ell + 1 )^2 - \frac{1}{4}}{r^2} + \left( \nu^2 -\frac{1}{4}\right) + \frac{9M}{4 r^{4}} - \frac{21Q^2}{r^{6}} \right],
\end{equation}
where
\begin{equation}
    \nu = \sqrt{4 +m^2} .
\end{equation}
To proceed with the semiclassical approximation, we define
\begin{equation}
    \ell + 1 := \nu \tilde{\ell}, \quad \omega = \nu \tilde{u},
\end{equation}
and take $\nu$ to be large. 

The potential splits as the sum of the two terms
\begin{equation}
    V_{0}=  \nu^2 f( 1 + \frac{\tilde{\ell}^2}{r^2}), \qquad V_{1} = f \left[ -\frac{1}{4r^2} - \frac{1}{4} + \frac{9}{4}  \frac{M}{r^{4}} -21 \frac{Q^2}{r^{6}}\right].
\end{equation}
Writing the wavefunction as $e^{\nu S}$ and expanding as
\begin{equation}
    S_{0}+ \nu^{-1} S_{1} + \nu^{-2}S_{2}, 
\end{equation}
we find three equations at each order of $\nu$
\begin{align}
    S_{0}' &= \pm \sqrt{V_{0}- \tilde{u}^2} := q, \\
    S_{1}' &= -\frac{1}{2} (\log S_{0}')' := -\frac{1}{2} \frac{q'}{q}, \\
    S_{2}'&= \frac{1}{2 S_{0}'} \left[ V_{1} - (S_{1}')^2 - S_{1}'' \right] = \frac{V_{1}}{2q} + \frac{1}{8} \frac{(q')^2}{q^3} + \frac{1}{4}\left( \frac{q'}{q^2}\right)'.
\end{align}
At each order in the mass, the equations above are solved up to an integration constant. This can be fixed by matching to the near-boundary expression at the corresponding order, as the radial equation is analytically solvable when the background is global AdS. 

The solutions to the first two equations are
\begin{equation}
    S_{0} = -\int_{\mathcal{C}} \frac{dr}{f(r)}  \sqrt{V_{0} - \tilde{u}^2}, \qquad S_{1} = -\frac{1}{2} \log \sqrt{V_{0}-\tilde{u}^2},
\end{equation}
where the contour $\mathcal{C}$ is chosen to go from $\infty$ around the turning point radius $r_{t}$ and then back to $\infty$. The two-point function in the geodesic approximation is obtained  by extracting the asymptotic normalization of the WKB function, after determining its relative normalization with the bulk wavefunction at the horizon. At the outer horizon, we have the ingoing boundary condition with a phase shift.  

The one-loop term has been written as a sum of a total derivative in $z$ and a local piece to be integrated. The boundary contribution vanishes using the same contour, and the solution to the one-loop order equation is
\begin{equation}
      S_{2}= \int_{\mathcal{C}} \frac{dr} {f} \left[  \frac{V_{1}}{2q} + \frac{f^2}{8} \frac{(\partial_{r}q)^2}{q^{3}}\right]
\end{equation}
which is finite at infinity and does not require further subtraction. 

\subsection{Analytic structure of quantum corrections}
\label{app:WKBbreakdown}

We now describe a systematic treatment of the analytic structure of the WKB approximation to the two-point function, incorporating higher order corrections. The leading order term is governed by the WKB phase
\begin{equation}
    Z = \int \frac{dr \sqrt{f( 1 - \frac{L^2}{r^2}) - E^2}}{f}. 
\end{equation}
The above object has branch points at the turning points. To understand its pole structure, we make use of the fact that 
\begin{equation}
    f = \frac{p_{6}}{r^4}, 
\end{equation}
where $p_{6}$ is a sixth order polynomial in $r$. Then, we see that
\begin{equation}
    Z = \int \frac{dr \sqrt{p_{6}(r^2 - L^2) - E^2 r^6}}{r^3 \frac{p_{6}}{r^4}}= \int dr \frac{r\sqrt{p_{6}(r^2 - L^2) - E^2 r^6}}{p_{6}}.
\end{equation}
The pole structure comes entirely from $p_{6}$, which has roots at the radii $\pm r_{+}, \pm r_{-}, \pm \pm i (1 + r_{+}^2 + r_{-}^2)$. 

The one-loop correction to the above picture is governed by, at leading order in energy, a term of the form
\begin{equation}
    Z_{1} = \int \frac{dr V_{1}}{f \sqrt{f(1- \frac{L^2}{r^2}) - E^2}}. 
\end{equation}
Using the following 
\begin{equation}
    \frac{V_{1}}{f} = \frac{\tilde{p}_{6}}{r^6}, 
\end{equation}
and the similar relation for $f$, we can simplify this to
\begin{equation}
   Z_{1} = \int dr \frac{\tilde{p}_{6}}{r^3 \sqrt{ p_{6}(r^2 - L^2) - E^2 r^6}} .
\end{equation}
Now, consider the contour $\mathcal{C}$ which starts from infinity, goes around the exterior turning point, and returns to infinity. The analytic continuation in $E$ that exchanges the exterior and interior turning points will then drag the defining contour and envelop the poles at the horizons. Thus, we expect that an analytic continuation of $Z$ in $E$ to pick up additional terms sensitive to the interior. 

To make this sharper, we will work with the strict large energy limit. In that case, the inner turning point $r_{t_{-}}$ scales as $\frac{1}{\sqrt{E}}$, which can be seen by finding a small $r$ root of the turning point equation. We will imagine that we have analytically continued the two-point function so as to exchange the exterior turning point with the interior one. Then, the one-loop term near the singularity goes as
\begin{equation}
    Z_{1} \sim_{E \to \infty} -\frac{i}{E r_{t_{-}}^{5}} = -i E^\frac{3}{2},
\end{equation}
which diverges.

It is also straightforward to obtain integral expressions for the higher order corrections, using the same method. In the same notation, we have 
\begin{equation}
    S_{3}' = - \frac{1}{2S_{0}'} \left[ S_{2}'' + 2 S_{1}' S_{2}'\right]. 
\end{equation}
The full expression of this $\frac{1}{\nu^2}$ correction to the exponent of the function is complicated, so we will choose the leading order in $u$ contribution as we did for $S_{2}$ to diagnose the change in the analytic structure. We have checked that the remaining terms are qualitatively similar.

We consider the leading order in energy term to $S_{3}$, which is given by
\begin{equation}
    Z_{3}= - \frac{1}{4} \int dr \frac{\partial_{r}V_{1}}{f ( 1 - \frac{L^2}{r^2}) - E^2}. 
\end{equation}
Since $V_{1} = f\frac{\tilde{p}_{6}}{r^6} = \frac{p_{6} \tilde{p}_{6}}{r^{10}}$, we can define $\partial_{r} V_{1} = \frac{h_{12}}{r^{11}}$ where $h_{12}$ is a $12$th order polynomial such that
\begin{equation}
    - \frac{1}{4} \int dr \frac{\partial_{r}V_{1}}{ f ( 1 - \frac{L^2}{r^2}) - E^2} = - \frac{1}{4}  \int dr \frac{h_{12}}{r^5 \left(p_{6} ( r^2 - L^2) - E^2 r^6 \right)}.
\end{equation}
Near the singularity and in the large $E$ limit, this scales as 
\begin{equation}
    S_{3} \sim_{E_{\to \infty}} E^{3}, 
\end{equation}
which diverges faster than the one-loop term.

\section{Steepest descent contour in rotating BTZ}
\label{app:contour}

\begin{figure}[t]
    \centering
    \includegraphics[scale=0.27]{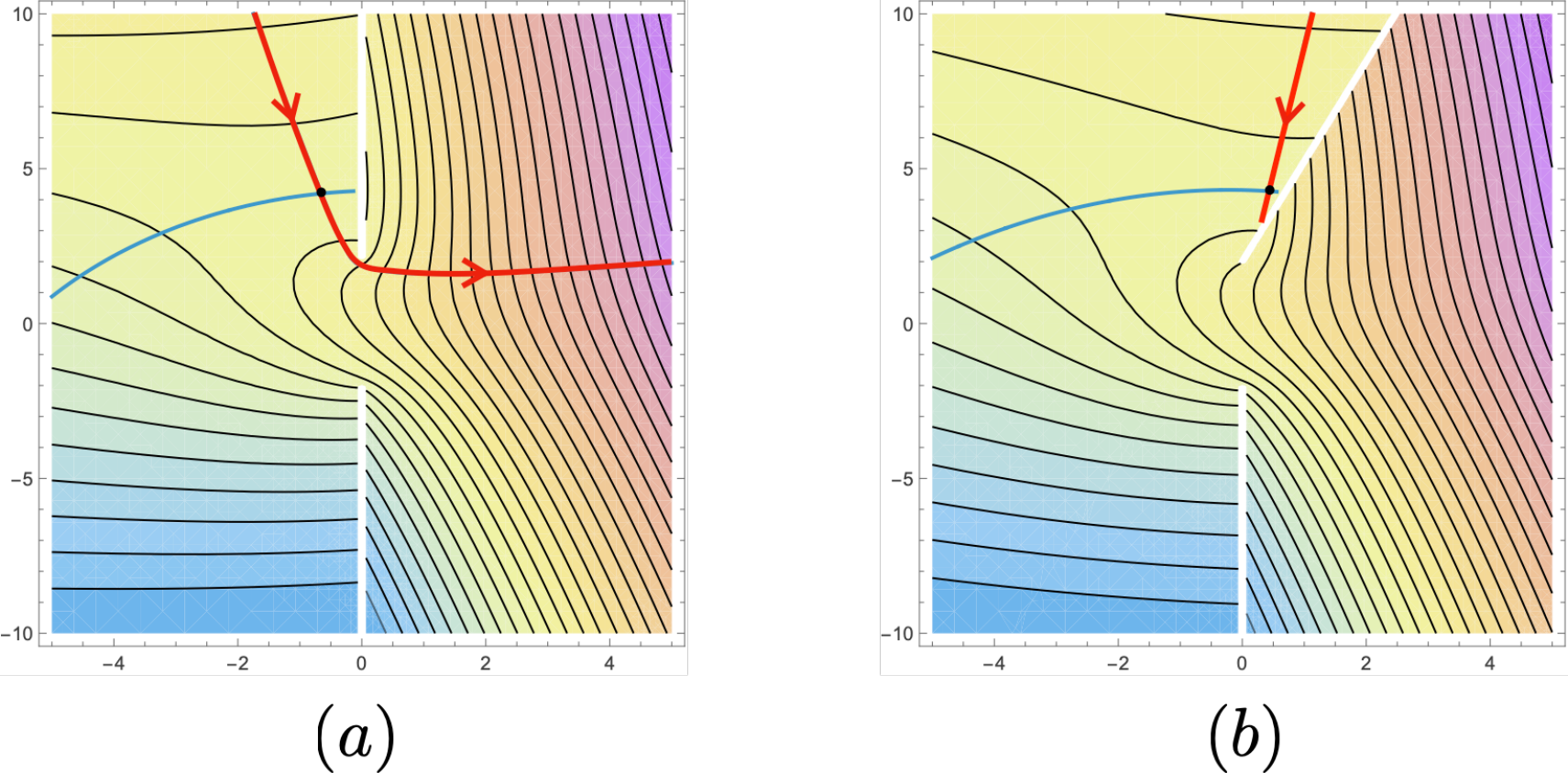}
    \caption{The path of steep descent (red contour) before and after the transition at $\text{Im}(y)=\beta$ in the $\bar{p}$ plane. (a) shows before the transition and (b) shows right after. In (b) we have slightly rotated the branch cut of $A_-$ in order to reveal the sunken saddle point.}
    \label{fig:Stokesp}
\end{figure}

In this appendix, we study in detail the steepest descent contour relevant for the analytic continuation outlined in the BTZ section. Without loss of generality, we focus on the holomorphic integral $G^+(y)$ of the Fourier transform in the position space two-point function $G^+(y,\bar{y})$:
\begin{equation}
\label{eq:GintBTZ}
    G^+(y) = \int d\bar{p} \exp\left(\frac{i\bar{p}y}{2}+m\bar{Z}(\bar{p})\right),
\end{equation}
where
\begin{equation}
    \bar{Z}(\bar{p}) = \frac{\beta}{4\pi i}(A_+\ln A_+-A_-\ln A_-)-\ln 2, \quad A_\pm \equiv \frac{\bar{p}}{m}\pm \frac{2\pi i}{\beta},
\end{equation}
and the default integration contour runs from $\bar{p}=-\infty$ to $p=+\infty$.
The function $\bar{Z}(\bar{p})$ has two log branch cuts, with branch points located at $\bar{p}=\pm2\pi im/\beta$.
The formula for the saddle point was given in \eqref{eq:BTZsaddle} which we reproduce here:
\begin{equation}
    \bar{p}^* = \frac{2\pi im}{\beta}\coth \frac{\pi y}{\beta}
\end{equation}
We are interested in the evolution of the steepest descent contour as we continue $y=y_0$ to $y=y_0+i\beta$. We will pick $y_0=i\tau_0+\epsilon$ with $0<\tau_0\le\beta/2$ and some $\epsilon>0$.
As we increase the imaginary part of $y$, the saddle point $\bar{p}^*$ rotates around the branch point $\bar{p}=2\pi im/\beta$ and sinks under the branch cut when $\text{Im}(y)>\beta$. Accompanied with this, the steepest descent path separates from the principal sheet and can no longer be deformed into the defining integration contour. This is illustrated in FIG.~\ref{fig:Stokesp}.

\begin{figure}[t]
    \centering
    \includegraphics[scale=.27]{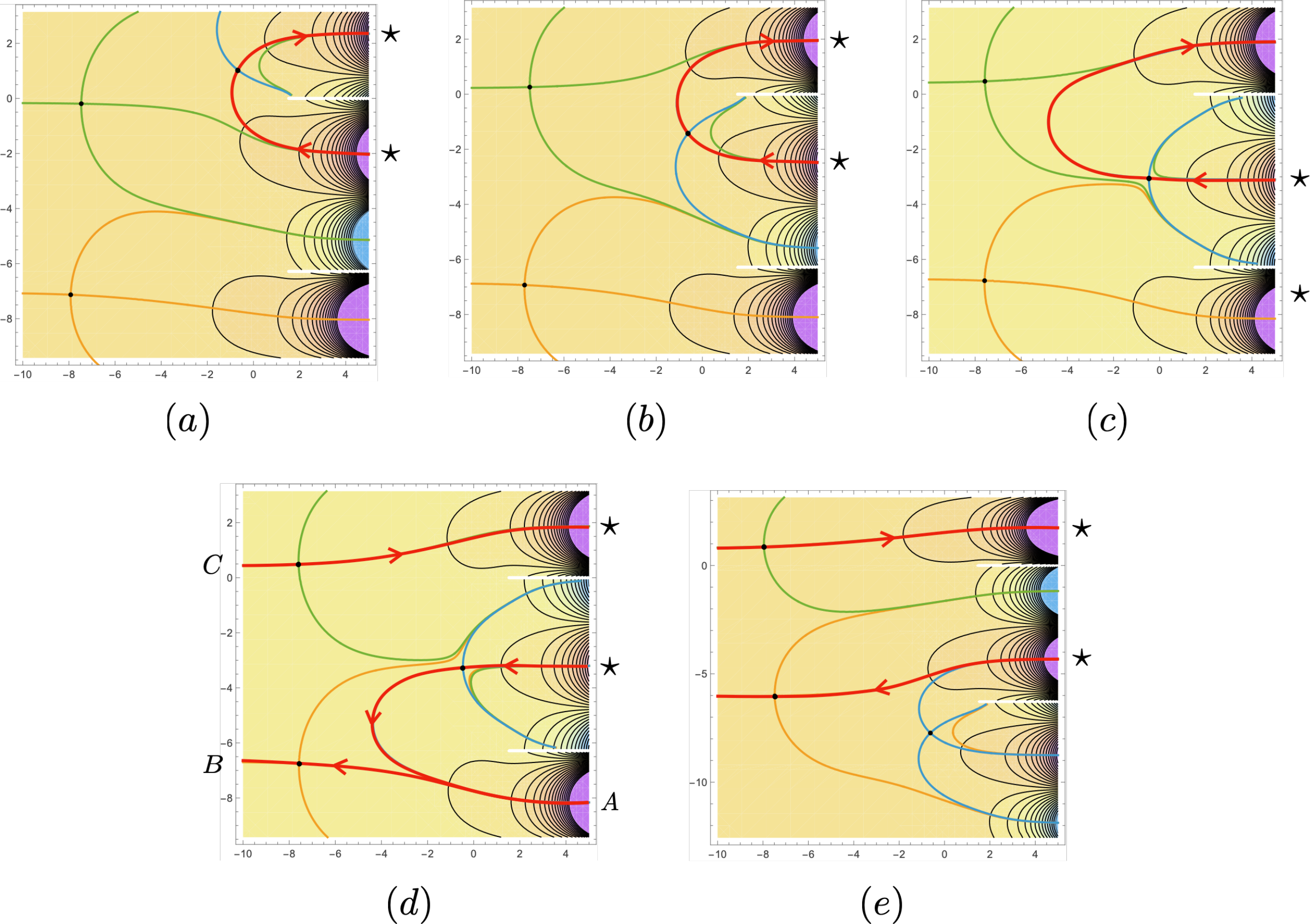}
    \caption{The evolution of the steepest descent contour (red) in the unwrapped coordinate. The asymptotic regions that defines the original contour are labeled by stars. The saddles are labeled by black dots and the path of steepest descent are labeled by lines with different colors originating from the saddle: The blue represents the principal saddle $\bar{q}^*$ and the green/orange represents the saddles $\bar{q}_1,\bar{q_2}$ arising from the inclusion of the Jacobian in the WKB action. The actual contour of the integral must be deformable to a union of steepest descent contours, with the starting point and the end point anchored to the star regions. The panels (a)-(e) shows different stages as we increase $\text{Im}(y)$ from $0$ to $3\beta/2$, during which the principal saddle $\bar{q}^*$ moves in the negative imaginary direction in the $\bar{q}$ coordinate. (a): $0<\text{Im}(y)<\beta/2$. There exists a simple contour which passes through $\bar{q}^*$ and connects the two asymptotic regions. (b): $\beta/2<\text{Im}(y)<\beta$. At $\text{Im}(y)=\beta/2$ a Stokes phenomenon happened between $\bar{q}^*$ and $\bar{q}_1$ and the ascending branch (blue contour) of the principal saddle jumped to the next sheet. This does not effect the red descending contour. (c): $\text{Im}(y)=\beta^-$. The situation right before the red contour shifts to the next sheet in the original $\bar{p}$ plane. This is the same as FIG.~\ref{fig:Stokesp}(a) but shown in the unwrapped coordinate. (d): $\text{Im}(y)=\beta^+$. The situation right after $\bar{q}^*$ crosses two Stokes lines and its contour shifts to the next sheet in the original $\bar{p}$ plane. The steepest descent contour $\star\to A\to B\to C\to \star$ now consists of three segments, where it passes through three saddles $\bar{q}^*\to\bar{q}_2\to\bar{q}_1$ in order. This is the same as FIG.~\ref{fig:Stokesp}(b) but shown in the unwrapped coordinate (The contours $A\to B$ and $C\to\star$ are not drawn in FIG.~\ref{fig:Stokesp}(b)). (e): $\text{Im}(y)>3\beta/2$. The ascending branch of $\bar{q}^*$ jumps again and now the steepest descent path that connects the starred asymptotic regions no longer passes through the principal saddle $\bar{q}^*$.}
    \label{fig:stokesq}
\end{figure}

The separation of the steepest descent contour can be traced back to a Stokes phenomenon associated with a second saddle. However, the saddle responsible for this ``coincides'' with the branch point at $A_-=0$ in the large $m$ limit and makes it difficult to analyze what actually happened. To resolve this issue, we introduce a uniformizing coordinate system in which the branch cut corresponding to $A_-$ is unwound. Defining the new coordinate $\bar{q}$ via $e^{\bar{q}} = A_-(\bar{p})$, we can write \eqref{eq:GintBTZ} as
\begin{equation}
\label{eq:GintBTZ2}
    G^+(y) = \int d\bar{q} \exp\left[\bar{q}+m\left(\frac{i(e^{\bar{q}}+2\pi i /\beta)y}{2}+Z(\bar{q})\right)\right].
\end{equation}
The $\bar{q}$ term in the exponential comes from the Jacobian of the coordinate transformation. Each sheet in the original $\bar{p}$ coordinate is mapped to an infinitely long horizontal strip with width $2\pi$ in the new coordinate system and moving along the imaginary direction in the $\bar{q}$ plane amounts to rotating around the $A_-$ branch point in the original coordinate. The asymptotic regions in the $\bar{p}$ plane are mapped to $+\infty$ in the $\bar{q}$ plane and the branch point $\bar{p}=2\pi im/\beta$ is mapped to $-\infty$. Note that although the $A_-$ branch cut is resolved in the new coordinate system, the $A_+$ branch cut still remains, manifesting itself as an infinite sequence of branch cuts separated by $2\pi i$; see, e.g. FIG.~\ref{fig:stokesq} for illustrations.

The term in parenthesis in \eqref{eq:GintBTZ2} is the same as that of \eqref{eq:GintBTZ}, but in the new $\bar{q}$ coordinate. Therefore, in the WKB limit $m\to\infty$ we can drop the Jacobian and the new saddle point in $\bar{q}$ plane is a simple coordinate reparametrization of the old one:
\begin{equation}
    \bar{q}^*=\ln(A_-(\bar{p}^*)) = \ln\left(\frac{4\pi i}{\beta}\frac{1}{e^{2\pi y/\beta}-1}\right)
\end{equation}
We will refer to $\bar{q}^*$ as the \emph{principal saddle} from now on.
Since in our setup we always have $\text{Re}(y)>0$, as one continues $y\to y+i\beta$, the terms in the parenthesis winds around the origin in clockwise direction and thus the saddle $\bar{q}^*$ moves in the negative imaginary direction in the $\bar{q}$ plane.

The presence of the Jacobian term in \eqref{eq:GintBTZ2} drastically changes the behavior of the WKB phase close to the original branch point of $A_-$.
In the $\bar{q}$ plane this corresponds to $\text{Re}(\bar{q})\to-\infty$, in which we can approximate $Z'(\bar{q})\sim \beta \bar{q}e^{\bar{q}}/4\pi i $ and thus saddle point condition becomes
\begin{equation}
\label{eq:Jsaddle}
   \frac{m \beta}{4\pi i}\bar{q}e^{\bar{q}} +1\approx0 \quad \Rightarrow \quad \bar{q}e^{\bar{q}} \approx \frac{4\pi}{im\beta}.
\end{equation}
This equation can have multiple solutions for $\text{Re}(\bar{q})\ll0$ when $\bar{q}e^{\bar{q}}\sim m^{-1}$. In particular, say if $\bar{q}_0$ is a solution for \eqref{eq:Jsaddle}, then $\bar{q}_n\equiv \bar{q}_0+2\pi n i$ also approximately satisfies \eqref{eq:Jsaddle} for $n\in \mathbb{Z}$. Therefore, we see that the inclusion of the Jacobian term in \eqref{eq:GintBTZ2} generates an infinite sequence of saddle points $\bar{q}_n$, roughly separated by integer multiples of $2\pi i$ with real part $\text{Re}(\bar{q}_n)\simeq W(4\pi/m\beta)$ where $W$ is the Lambert W function. In the limit of large $m$ these additional saddles approach the branch point $\bar{q}=-\infty$.

These new saddles $\bar{q}_n$ can be used to generate additional steepest descent contours that the defining integration contour can deform into when $\text{Im}(y)>\beta$. This is demonstrated in FIG.~\ref{fig:stokesq}: For $\text{Im}(y)<\beta$, the defining contour can always be deformed into the steepest descent contour of the principal saddle (panels (a)-(c)). When $\text{Im}(y)>\beta$, the defining contour can be deformed into a contour that passes through the principal saddle $\bar{q}^*$ and two additional saddles $\bar{q}_1,\bar{q}_2$ (panel (d)). In the large $m$ limit, the WKB action at $\bar{q}_n$ is approximately given by
\begin{equation}
    -\left(\frac{\pi y}{\beta}-\ln \frac{4\pi i}{\beta}\right)+O\left(\frac{\bar{q}_n}{m}\right),
\end{equation}
which does not depend on $\bar{q}_n$ to the leading order in $m$. Since the steepest descent path goes through $\bar{q}_1,\bar{q}_2$ in the opposite direction, their contributions cancel and only the contribution of $\bar{q}^*$ remains, which gives the desired $\pi i$ shift of the Lorentzian geodesic excursion. This however stops working when $\text{Im}(y)>3\beta/2$: The asymptotic regions of the principal saddle jump again and now the relevant steepest descent contour only consists of the saddles $\bar{q}_1$ and $\bar{q}_2$ (panel (e)).

\section{5d AdS-Kerr}
\label{app:Kerr}
The metric of a 5d AdS-Kerr (AdS-Myers-Perry) black hole \footnote{We work with the Lorentzian metric in this appendix. The Euclidean form of the metric and equation of motions can be obtained by a simple continuation $t\to i\tau$ and $L\to iL$.} in a Boyer-Lindquist-like coordinate which is static at infinity is given by \cite{Myers:1986un,Hawking:1998kw,Grunau:2017uzf}:
\begin{align}
\begin{split}
ds^2 =& -\frac{\Delta_\theta(1+r^2)dt^2}{\Xi_a\Xi_b}+\frac{2M}{\rho^2}\left[\frac{\Delta_\theta dt}{\Xi_a\Xi_b}-\frac{a\sin^2\theta}{\Xi_a}d\phi-\frac{b\cos^2\theta}{\Xi_b}d\psi\right]^2  \\
&+\frac{r^2+a^2}{\Xi_a^2}\sin^2\theta d\phi^2 +\frac{r^2+b^2}{\Xi_b^2}\cos^2\theta d\psi^2 + \frac{\rho^2r^2}{\Delta_r}dr^2 + \frac{\rho^2}{\Delta_\theta}d\theta^2,
\end{split}
\end{align}
where
\begin{align}
\begin{split}
\rho^2 &= r^2 + a^2\cos^2\theta+b^2\sin^2\theta, \quad  \Delta_r = (r^2+a^2)(r^2+b^2)(1+r^2)-2Mr^2,  \\
\Delta_\theta &= 1-a^2\cos^2\theta-b^2\sin^2\theta,\quad \Xi_a = 1-a^2, \quad \Xi_b = 1-b^2,
\end{split}
\end{align}
 The parameters $M,a,b$ are related to the mass and two angular momenta of the black hole. We work in the units where the AdS radius $\ell_{AdS}=1$.
 For constant $t$ and $r$, the metric describes a three-sphere in an oblique Hopf parameterization. For the purpose of this work we will only consider the case where the two angular momenta are equal, i.e. $a=b$.

 The horizons are given by the solutions of $\Delta_r=0$. For the spacetime to be devoid of naked singularity, we need $\Delta_r$ to have two positive solutions. Note that the coordinate $r$ does not cover the entire spacetime. We can cover the entire spacetime by passing to the $\rho$ coordinate.
 The curvature singularity is located at $\rho=0$, or $r^2=-a^2$ in the case of equal angular momenta.

 The geodesics are determined by the three conserved charges associated to the Killing vectors $(-\partial_t,\partial_\phi,\partial_\psi)$ which we denote $(E,L_\phi,L_\psi)$.
 The geodesic equations are separable with the help of a fourth constant of motion known as Carter's constant $K$ \cite{Carter:1968rr}. The explicit form of the geodesic equations can be found in \cite{Delsate:2015ina,Grunau:2017uzf} and we report the equal angular momenta case here:
 \begin{align}
 \rho^6\dot{\rho}^2 = R(\rho),\quad  \rho^4\dot{\theta}^2 = \Theta(\theta), \quad \rho^2\dot{t} = T(\rho),
 \end{align}
 where
 \begin{align}
 \begin{split}
 R(\rho) =& -\Delta_x(K+m^2(\rho^2-a^2))-E^2(-2Ma^2(2\rho^2-a^2)+\rho^4(\rho^2(1-2a^2)-a^2+a^4)) \\
  &+2Ma^2(L_\phi^2+L_\psi^2)(\rho^2-a^2+1) -4M(E(L_\phi+L_\psi)a\rho^2+L_\phi L_\psi a^2 (\rho^2-a^2+1)),
  \end{split} \\
  \Theta(\theta) =& \frac{K}{1-a^2} + E^2a^2(1-a^2)-\left[\frac{m^2a^2}{1-a^2}+\frac{L_\phi^2}{\sin^2\theta}+\frac{L_\psi^2}{\cos^2\theta}\right], \\
  T(\rho) =& \frac{1}{\Delta_x} \bigg[ E\big(2Ma^2(2\rho^2-a^2)+\rho^4(\rho^2(1-2a^2)-a^2+a^4)\big)
  -2Ma\rho^2(L_\phi+L_\psi)\bigg] + a^2E,
 \end{align}
 where $m^2 \in\{0,1,-1\}$ indicates the geodesic is null/timelike/spacelike.
 The turning points are given by the condition $R(\rho_t)=0$.

 Focusing on the null case, we see that the following combination of charges
 \begin{equation}
    m^2=0,\quad  L_\phi=L\sin^2\chi, \quad L_\psi=L\cos^2\chi, \quad K = L^2-a^2(L^2+E^2) 
 \end{equation}
 leads to $\Theta(\chi)=0$ (so that the motion in $\theta$ direction is trivial) and
\begin{align*} 
R(\rho)=\rho^{2}\left[(E^2-L^2)(1-a^2)\rho^4-L^2(1-a^2)^2\rho^2+2M(aE-L)^2\right].
\end{align*}
We see that $R$ has a double root at $\rho=0$.
However, recall that since $\dot{\rho}^2 = \rho^{-6}R(\rho)$, this does not mean that it corresponds to an actual turning point.
Rather, it represents a geodesic that starts or terminates at $\rho=0$.
The actual turning points are given by the solution of the quadratic equation in the square bracket.
For the particle to be able to escape to infinity we need to make sure there is no other turning points for $\rho^2>0$.
We can guarantee this if the discriminant $\Delta$ of the equation is less than zero, i.e.
\begin{equation}
\label{discriminant}
\Delta \propto L^2(1-a^2)^3- 8M\left(1-\frac{L^2}{E^2}\right)\left(aE-L\right)^2 \le 0 .
\end{equation}
In particular this is always true for $L=0$.

To see how this null geodesic can be approximated by timelike ones we need to examine the geodesic equation for nonzero $m^2$.
Consider the following combination of charges
\begin{equation}
    m^2=1, \quad L_\phi=L\sin^2\chi, \quad L_\psi=L\cos^2\chi,\quad K=L^2-a^2(L^2+E^2)+a^2(m^2+k),
\end{equation}
where $k$ is some order one number which remains unfixed for now. We will consider the behavior of this geodesic for large $E$ while keeping the ratio $L/E$ fixed.
With this choice we have
\begin{equation}
    R(\rho) = R_8\rho^8 + R_6\rho^6 + R_4\rho^4 + R_2 \rho^2 + R_0,
\end{equation}
with the individual coefficients given by
 \begin{align}
  R_8 &= -1, \\
  R_6 &= (1-a^2)(E^2-L^2)-a^2(k-1)-1, \\
  R_4 &= -(1-a^2)^2L^2-a^2(1-a^2)k+2M, \\
  R_2 &= 2M((aE-L)^2+a^2(k-1)), \\
  R_0 &= -2a^4Mk.
 \end{align}
For $k=0$, there is a root at $\rho=0$. One can show that additionally there is only one positive root at
\begin{equation}
    \rho_{t_+} \approx \sqrt{(1-a^2)(E^2-L^2)}
\end{equation}
in the regime where $E,L\gg 1$ provided \eqref{discriminant} is satisfied.
This geodesic thus represents a massive particle that falls from $\rho_{t_+}$ into the singularity.
By turning on a small but nonzero $k>0$, the particle develops a turning point near the singularity at
\begin{equation}
   \rho_{t_-} \approx 2a^4Mk,
\end{equation}
while the other features of the effective potential remain unchanged.
The particle should now be viewed as following a geodesic that connects two different exterior regions in the maximally extended Penrose diagram. As one takes $E\to \infty$, this geodesic approaches the union of a pair null geodesics we constructed previously, connected at the singularity $\rho=0$.

Note that with this choice the $\Theta$ potential is no longer zero, instead we have
\begin{equation}
\Theta(\theta) = a^2(1-a^2)k-\frac{(1-a^2)^2L^2}{4\sin^2\theta\cos^2\theta}(\cos(2\chi)-\cos(2\theta))^2.
\end{equation}
For $k=0$, the particle is confined to the hyperplane $\theta=\chi$ as we have seen above.
For $k>0$, the particle oscillates between the two zeros of $\Theta(\theta)$, which can be approximated in the small $k/L$ limit to be
\begin{equation}
    \theta_{t_\pm} = \chi \pm \frac{ak}{2L\sqrt{1-a^2}}.
\end{equation}

To summarize, we have shown a null geodesic that bounces between two asymptotic regions can be approximated by a series of timelike geodesics with 
\begin{equation}
    m^2=1,\quad L_\phi=L\sin^2\chi, \quad L_\psi=L\cos^2\chi,\quad K=L^2-a^2(L^2+E^2)+a^2(Q+k),
\end{equation}
in the limit
\begin{equation}
    E\to \infty, \quad  \frac{L}{E}\sim O(1),\quad k\to 0^+
\end{equation}
provided \eqref{discriminant} is satisfied.

\end{document}